\documentclass{aa}  
\usepackage[varg]{txfonts}
\usepackage{graphicx}
\usepackage[hidelinks,colorlinks=true,breaklinks,allcolors=blue]{hyperref} 
\usepackage{natbib}
\bibpunct{(}{)}{;}{a}{}{,} 

\usepackage{orcidlink}

\defcitealias{2023Zewdie}{Z23}

\begin{document} 
\title{Overdensity of Lyman-Break Galaxy Candidates Around Hot Dust-Obscured Galaxies}

\titlerunning{ An Overdensity of LBGs Around Hot DOGs}

\author{Dejene Zewdie\inst{1, 2}\orcidlink{0000-0003-4293-7507}
          \and
Roberto J. Assef\inst{1}\orcidlink{0000-0002-9508-3667}
\and
Trystan Lambert\inst{1,3}\orcidlink{0000-0001-6263-0970}
\and
Chiara Mazzucchelli\inst{1}\orcidlink{0000-0002-5941-5214}
\and
S Ilani Loubser\inst{2}\orcidlink{0000-0002-3937-7126}
\and
Manuel Aravena\inst{1}\orcidlink{0000-0002-6290-3198}
\and
Jorge González-López\inst{4, 5}\orcidlink{0000-0003-3926-1411} 
\and
Hyunsung D.~Jun\inst{6,7}\orcidlink{0000-0003-1470-5901}
\and
Chao-Wei Tsai\inst{8,9,10}\orcidlink{0000-0002-9390-9672}
\and
Daniel Stern\inst{11}\orcidlink{0000-0003-2686-9241}
\and 
Guodong Li\inst{8,10}\orcidlink{0000-0003-4007-5771}
\and
Rom\'an Fern\'andez Aranda\inst{12,13}\orcidlink{0000-0002-7714-688X}
\and
Tanio D\'iaz-Santos\inst{13,14}\orcidlink{0000-0003-0699-6083}
\and
Peter R. M. Eisenhardt\inst{11} 
\and
Andrey Vayner\inst{15}\orcidlink{0000-0002-0710-3729}
\and 
Lee R. Martin\inst{16}\orcidlink{0000-0002-4950-7940}
\and 
Andrew W. Blain\inst{16}\orcidlink{0000-0001-7489-5167}
\and
Jingwen Wu\inst{8, 10}\orcidlink{0000-0001-7808-3756}
          }

\institute{Instituto de Estudios Astrof\'isicos, Facultad de Ingenier\'ia y Ciencias, Universidad Diego Portales, Av. Ej\'ercito Libertador 441, Santiago, Chile
\and 
Centre for Space Research, North-West University, Potchefstroom 2520, South Africa
\and
ICRAR, The University of Western Australia, 35 Stirling Highway, Crawley, WA 6009, Australia
\and
Instituto de Astrof\'isica, Facultad de F\'isica, 
  Pontiﬁcia Universidad Cat\'olica de Chile, Santiago 7820436, Chile
  \and
  Las Campanas Observatory, Carnegie Institution of Washington, 
  Ra\'ul Bitr\'an 1200, La Serena, Chile
\and
Department of Physics, Northwestern College, 101 7th St SW, Orange City, IA 51041, USA
\and
School of Physics, Korea Institute for Advanced Study, 85 Hoegiro, Dongdaemun-gu, Seoul 02455, Republic of Korea
\and 
National Astronomical Observatories, Chinese Academy of Sciences, 20A Datun Road, Beijing 100101, China
\and
Institute for Frontiers in Astronomy and Astrophysics, Beijing Normal University, Beijing 102206, China    
\and 
University of Chinese Academy of Sciences, Beijing 100049, China   
\and 
Jet Propulsion Laboratory, California Institute of Technology, 4800 Oak Grove Drive, Pasadena, CA 91109, USA
\and
Department of Physics, University of Crete, 70013 Heraklion, Greece
\and
Institute of Astrophysics, Foundation for Research and Technology–Hellas (FORTH), Heraklion, GR-70013, Greece
\and
School of Sciences, European University Cyprus, Diogenes street, Engomi, 1516 Nicosia, Cyprus
\and
IPAC, California Institute of Technology, 1200 E. California Boulevard, Pasadena, 91125, CA, USA
\and
Physics \& Astronomy, University of Leicester, 1 University Road, Leicester LE1 7RH, UK
}             

\date{Received July 08, 2024; accepted December 04, 2024}

\abstract{Hot dust-obscured galaxies (Hot DOGs), are a family of hyper-luminous, heavily obscured quasars. A number of studies have shown that these objects reside in significantly overdense regions of the Universe based on the identification of companions at optical through far-IR wavelengths. Here we present further characterization of their environments by studying the surface density of Lyman break galaxy (LBG) candidates in the vicinity of three Hot DOGs. For two of them, WISE J041010.60–091305.2 (W0410--0913) at $z=3.631$ and WISE J083153.25+014010.8 (W0831+0140) at $z=3.912$, we identify the candidate LBG companions using deep observations obtained with Baade/IMACS. For the third, WISE J224607.56--052634.9 (W2246--0526) at $z=4.601$, we re-analyse previously published data obtained with Gemini-S/GMOS-S. We optimise the LBG photometric selection criteria at the redshift of each target using the COSMOS2020 catalog. When comparing the density of LBG candidates found in the vicinity of these Hot DOGs with that in the COSMOS2020 catalog, we find overdensities of $\delta=1.83\pm 0.08$ ($\delta' = 7.49\pm 0.68$), $\delta=4.67\pm 0.21$ ($\delta' = 29.17\pm 2.21$), and $\delta = 2.36\pm 0.25$ ($\delta' = 11.60\pm 1.96$) around W0410--0913, W0831+0140, and W2246--0526, respectively, without (with) contamination correction. Additionally, we find that the overdensities are centrally concentrated around each Hot DOG. Our analysis also reveals that the overdensity of the fields surrounding  W0410--0913 and W0831+0140 declines steeply beyond physical scales of $\sim$2~Mpc. If these overdensities evolve to clusters by $z=0$, these results suggest that the Hot DOG may correspond to the early formation stages of the brightest cluster galaxy. We were unable to determine if this is also the case for W2246--0526 due to the smaller field of view of the GMOS-S observations. Our results imply that Hot DOGs may be excellent tracers of protoclusters. }


  \maketitle

\section{Introduction}

The hierarchical assembly of galaxies implies that the environment in which galaxies are born and live can play a fundamental role in driving their evolution \citep[e.g.,][]{2007Li, 2019Dayal}. This merging process may trigger active galactic nuclei (AGNs) activity, driving the growth of the supermassive black holes (SMBHs) at their centers. Since we know that SMBHs assemble the majority of their mass through gas accretion during AGN phases \citep{1982Soltan, 2020Inayoshi}, it is likely then that the most luminous quasars live in overdense regions of the Universe. In particular, the existence of SMBHs with $\rm >10^{8-9} M_{\odot}$ in the early Universe \citep[$z\gtrsim 6$; e.g.,][for a recent review]{2018Banados, 2021Wang, 2023Fan}, implies that these sources must be fed by large amounts of gas, and that they must live in the densest regions at that time \citep{2009Overzier, 2012Angulo}. The observational evidence, however, seems contentious.

Spanning over two decades, the exploration of overdensities around luminous high-redshift quasars and radio galaxies has notably deepened our comprehension of their environment throughout cosmic time \citep[e.g.,][]{2006Zheng, 2009Kim, 2010Utsumi, 2013Husband, 2014Morselli, 2017GarcaVergara,2019Garcia, 2017Mazzucchelli, 2018Uchiyama, 2020Mignoli, 2024Lambert}. Spectroscopically identifying companion galaxies around quasars and radio galaxies is difficult due to the inherent faintness of the surrounding galaxies, so various methods based purely on photometric observations have been devised to trace overdense regions. The most common of these include identification of either Lyman-break galaxies \citep[LBGs; e.g.,][]{2003Steidel, 2004VOuchi, 2006Yoshida,  2013Husband, 2014Morselli, 2017GarcaVergara} through broad-band optical colours, or Lyman-alpha emitters \citep[LAEs, e.g.,][]{2007Kashikawa, 2019Garcia} via a combination of broad- and narrow-band observations.

 Some studies find overdensities around high-redshift radio galaxies \citep[$1<z<5$; e.g.,][]{2002Venemans, 2004Miley, 2006Intema, 2012Mayo, 2004Venemans, 2007Venemans, 2020Bosman} and high-redshift quasars \citep[$z \gtrsim 4$; e.g.,][]{2006Zheng, 2014Morselli, 2017Balmaverde, 2017GarcaVergara}. Others find a mix of overdensity and underdensity of LBGs \citep[$z\geq 6$; e.g.,][]{2009Kim, 2018Ota, 2023Champagne}. There are also studies that find no overdensity of LAEs/LBGs around high-redshift quasars \citep[$z>5.5$; e.g.,][]{2013Banados, 2017Mazzucchelli, 2018Uchiyama}. Recently, \cite{2024Lambert} found evidence that the quasar itself may hinder star-formation in its vicinity, suggesting that the use of LAEs as overdensity tracers should be taken with caution. Even under such potential caveats, these overdense environments provide a unique opportunity to understand the formation of large-scale structures, such as protoclusters in the early Universe.



Hot dust-obscured galaxies \citep[Hot DOGs;][]{2012Eisenhardt, 2012Wu}, discovered through the Wide-field Infrared Survey Explorer \citep[WISE;][]{2010Wright} mission, are among some of the most luminous and rare populations of quasars. These objects are powered by intense accretion onto SMBHs, heavily buried under enormous amounts of gas and dust \citep{2014Stern, 2015Tsai, 2018Tsai, 2015Assef}. Recently, \cite{2024Li} estimated the black hole masses of Hot DOGs using the broad C{\sc iv} and Mg{\sc ii} lines and found that they range from $10^{8.7}$ to $10^{10} M_\odot$.  Hot DOGs have extreme bolometric luminosities, $\rm L_{bol} > 10^{13}L_{\odot}$, with some exceeding $\rm L_{bol} > 10^{14}L_{\odot}$ \citep{2015Tsai}. These objects may play a significant role in the evolution of their host galaxies by inducing substantial gas outflows \citep{2016Tanio, 2020Jun, 2020Finnerty}. 

Previous studies of the environments of Hot DOGs have found that they inhabit densely populated regions \citep{2014Jones, 2017Jones, 2015Assef, 2017Fan, 2023Zewdie}. In particular, \cite{2015Assef} studied a large number of Hot DOGs through Spitzer/IRAC imaging and revealed that these objects statistically exist in dense environments similar to radio-loud AGN \citep[e.g.,][]{2013Wylezalek, 2018Noirot}. \cite{2014Jones, 2017Jones} explored the overdensities of submillimeter galaxies (SMGs) and mid-IR Spitzer-selected galaxies situated in proximity to Hot DOGs. Their findings also suggest that Hot DOGs could potentially reside in overdense environments. 
 
\cite{2022Ginolfi} studied the environment of WISE J041010.60–091305.2 (W0410--0913, $z=3.361$) using VLT/MUSE observations and found a significant overdensity of LAEs around this Hot DOG ($\delta = 14^{+16}_{-8}$, where \rm $\delta =\frac{N_{F}}{N_{E}}$, $N_{F}$ is the number of LAE/LBG candidates in targeted field, and $N_{E}$ is the number of LAEs/LBGs in the blank field, normalized to the area of the targeted field). \cite{2022Luo}  found double the surface density of distant red galaxies around W1835+4355 at $z=2.3$ compared to the field.  Recently, \citet{2023Zewdie} (hereafter \citetalias{2023Zewdie}) studied the environment of the most luminous known Hot DOG, WISE J224607.56--052634.9 (W2246--0526), at $z = 4.601$, using deep Gemini Multi-Object Spectrographs South (GMOS-S) imaging in the \textit{r}-, \textit{i}-, and \textit{z}-bands. They revealed a large overdensity ($\delta \sim 6$) of LBGs within 1.4 Mpc of the Hot DOG, suggesting it lives in an early stage proto-cluster. However, they did not observe a radial profile overdensity of LBGs around the Hot DOG. They interpreted this as implying that either the Hot DOG is not at the center, that the structure is too young to have a clear center even if the Hot DOG becomes the brightest cluster galaxy (BCG), or that observational effects from the selection canceled out the radial profile. Further indications of the overdense environment around this source have also been found by deep ALMA observations as \cite{2016Tanio, 2018Tanio} revealed the presence of companions around W2246--0526 linked by dust-streams up to 30~kpc from the Hot DOG. Their findings suggest that this system is a triple merger, in a locally dense environment.

In this work, we study the environment of three Hot DOGs; W0410--0913, WISE J083153.25+014010.8 (W0831+0140), and W2246--0526, through the identification of LBG companions. We use optimised selection functions based on the observations, redshifts, and classifications of the Cosmic Evolution Survey \citep[COSMOS;][]{2007Scoville, 2022Weaver} catalog. Specifically, we investigate LBG candidates around W0410--0913 ($z=3.631$) and W0831+0140 ($z=3.912$) selected as \textit{g}-band dropouts in imaging obtained with the Inamori-Magellan Areal Camera and Spectrograph (IMACS) at the Magellan Baade Telescope. Additionally, we re-analyse the data from \citetalias{2023Zewdie} for W2246--0526 using the same technique to optimise the selection function used for the other two fields. This paper is organized as follows: In Section \ref{ODR}, we discuss our IMACS observations, data reduction, photometric measurements, and the COSMOS2020 catalog. In Section \ref{optimazation}, we present the optimisation of the selection function based on the colour selection criteria. In Section \ref{sec_LBGc}, we discuss the study of the colour and spatial distribution of the LBG candidates and we compare our results with those for other Hot DOG and quasar environments presented in the literature. Our conclusions are presented in Section \ref{conclusion}. Throughout this paper, all magnitudes are given in the AB system. We assume a flat $\Lambda$CDM cosmology with $\rm H_0 = 70 ~km s^{-1} Mpc^{-1}$ and $\Omega_M = 0.3$. 


\section{Observations and data reduction} \label{ODR}
\subsection{Magellan/IMACS observations} \label{IMACS_data}

We used the IMACS instrument with the f/4 camera on the Magellan Baade telescope to obtain deep images in the \textit{g}-, \textit{r}-, and \textit{i}-bands of the fields around W0410--0913 and W0831+0140 on the night of UT2019-11-25 (P.I.: R.J. Assef). The average seeing was 1.02\arcsec, 1.07\arcsec, and 1.08\arcsec\, for the W0410--0913 observations in the \textit{g}-, \textit{r}-, and \textit{i}-bands, respectively, with airmass ranging from 1.07 to 1.15. For the W0831+0140 observations, the mean seeing was 0.81\arcsec, 1.10\arcsec, and 1.01\arcsec\ in the \textit{r}-, \textit{g}-, and \textit{i}-bands, respectively, with the airmass ranging from 1.17 to 1.54. All images have a pixel scale of 0.22\arcsec pix$^{-1}$ and a field of view (FoV) of 15.4\arcmin $\times$ 15.4\arcmin. A dithering pattern with offsets between 20\arcsec\ and 60\arcsec\ in R.A. and Dec. was applied to minimize the effect of the chip gaps in the final stacked image. The details of the observations are summarized in Table \ref{observation}. 


\begin{table*}
	\centering
\caption[]{Summary of IMACS and GMOS-S observations used in this work. All data were acquired during the night of UT 2019-11-25 or both IMACS observations and nights of UT 2017-09-16/23/27 for W2246--0526 GMOS-S observations (\citetalias{2023Zewdie}).}
\label{observation}
\begin{tabular}{c c c c c c cl}
	\hline
 	\text{\rm Field} & Instrument &\rm	Filters  & \text{\rm Total Exposure } &  \text{\rm Mean }& \text{\rm Mean }   & \text{Depth of the Stack Image}   \\
 &  &  & Time  & Seeing  & Airmass    &5 $\sigma$ /3 $\sigma$ /1 $\sigma$ \\
 	\hline
 	\noalign{\smallskip}
 	&&\textit{g}    & 4900s & 1.02\arcsec  & 1.15    & 26.70/27.25/28.43\\
W0410$-$0913 & IMACS & \textit{r} & 2100s & 1.07\arcsec &  1.08   & 25.83/26.39/27.56 \\
  & & \textit{i} & 1600s&  1.08\arcsec  & 1.07   & 25.13/25.69/26.88\\
\hline
 	\noalign{\smallskip}
 	&&\textit{g}   &  3600s &1.1\arcsec  & 1.30    & 26.42/26.96/28.10\\ 
W0831$+$0140  & IMACS & \textit{r}     & 1850s &  0.81\arcsec &  1.17  & 25.47/26.02/27.18 \\
  & &\textit{i}  & 1200s & 1.01\arcsec  & 1.54   & 25.08/25.63/26.83\\
\hline

 	\noalign{\smallskip}
& &\textit{r}   &  7500s &0.58\arcsec  & 1.1    & 27.29/27.84/29.00\\ 
W2246--0526 & GMOS-S & \textit{i}     & 7200s &  0.81\arcsec &  1.3  & 26.58/27.16/28.34 \\
 & & \textit{z}  & 6200s & 1.01\arcsec  & 1.2   & 26.02/26.57/27.77\\
\hline
\end{tabular}
\end{table*}

\subsection{IMACS data reduction}\label{imacs_reduction}

We reduced the IMACS observations by first applying bias and flat field corrections using {\tt{Theli}}\footnote{\url{https://github.com/schirmermischa/THELI}} \citep{2005Erben, 2013Schirmer}. The raw IMACS images do not come with World Coordinate System (WCS) information, so we developed a {\tt{python}} package\footnote{\url{https://github.com/TrystanScottLambert/imacs_wcs}} to calibrate the image astrometry. We first made an initial guess based on the telescope information in the image headers. We then cross-matched sources in our images with objects in Gaia DR3 \citep{2023Gaia} using the {\tt{astroquery}} package. The pixel to world relation was then used to define an accurate WCS object for each image. We found that the initial guess was off by about 20\arcsec. Finally, we used {\tt{Swarp}}\footnote{\url{https://www.astromatic.net/software/swarp/}} to coadd our images \citep{2002Bertin}. 

\begin{figure*}
\includegraphics[scale=0.45]{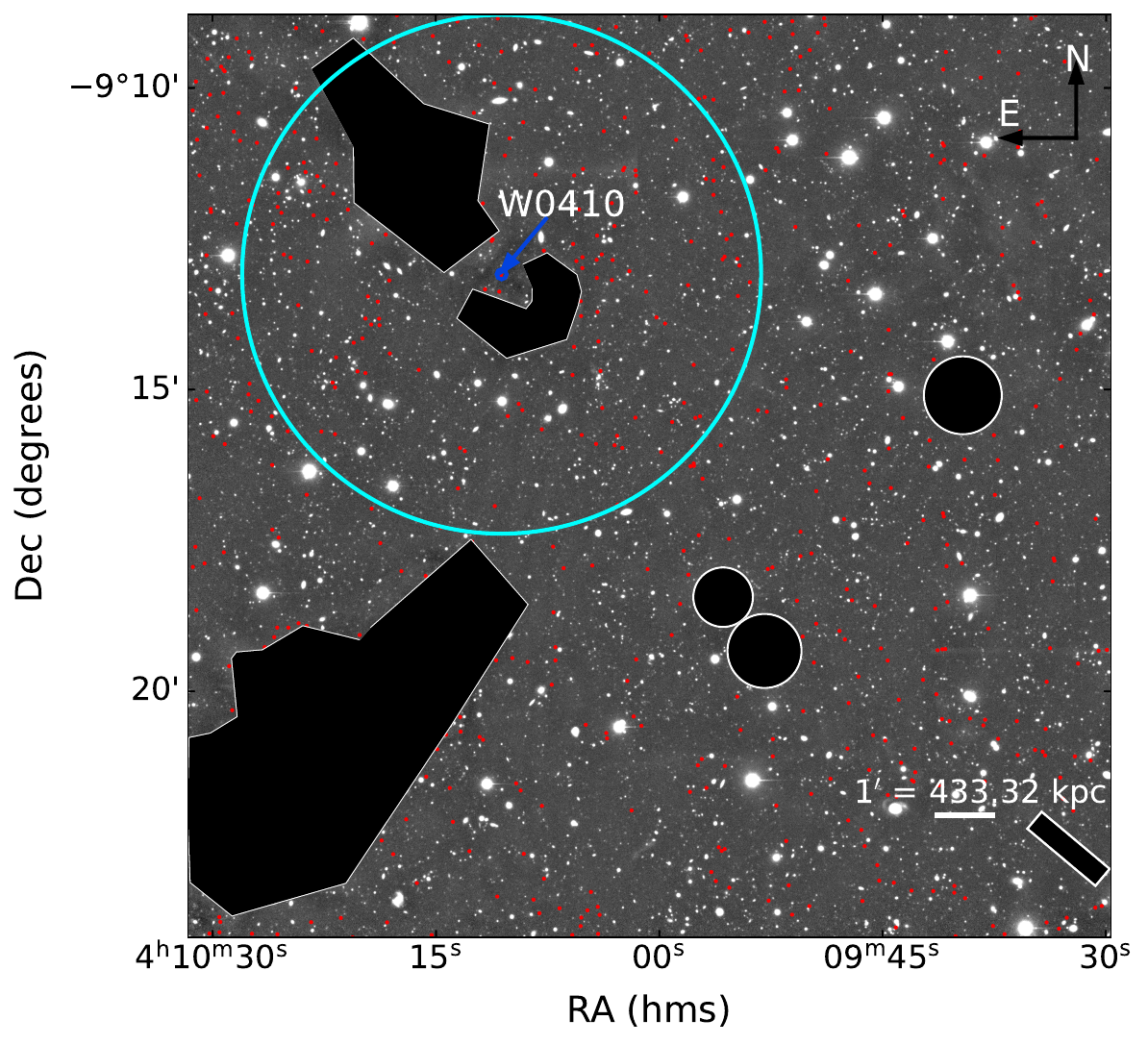} \includegraphics[scale=0.45]{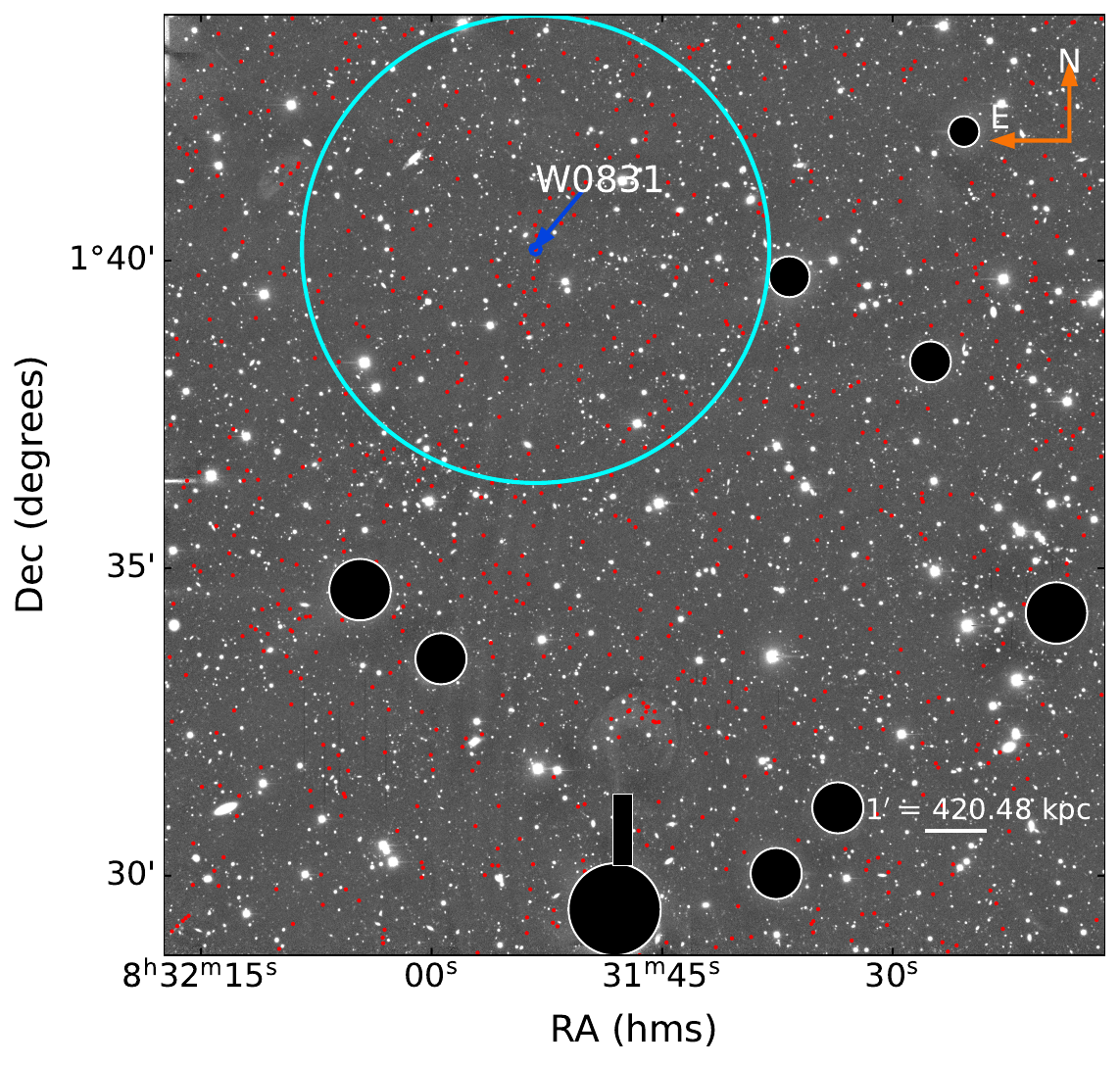}
\caption{\textit{r}-band image of W0410--0913 (left-panel) and W0831+0140 (right-panel). In both panels, black circles and polygons show the masked area that we did not used for our LBG selection (see Section \ref{IMACS_data} for details). The blue circles and arrows indicate the Hot DOGs positions, and the cyan circle represents the largest area, centered on the Hot DOG, within the image bounds. The red circles are candidate LBG companions.}
\label{ds9_image}
\end{figure*}

 \begin{figure}
\includegraphics[scale=0.31]{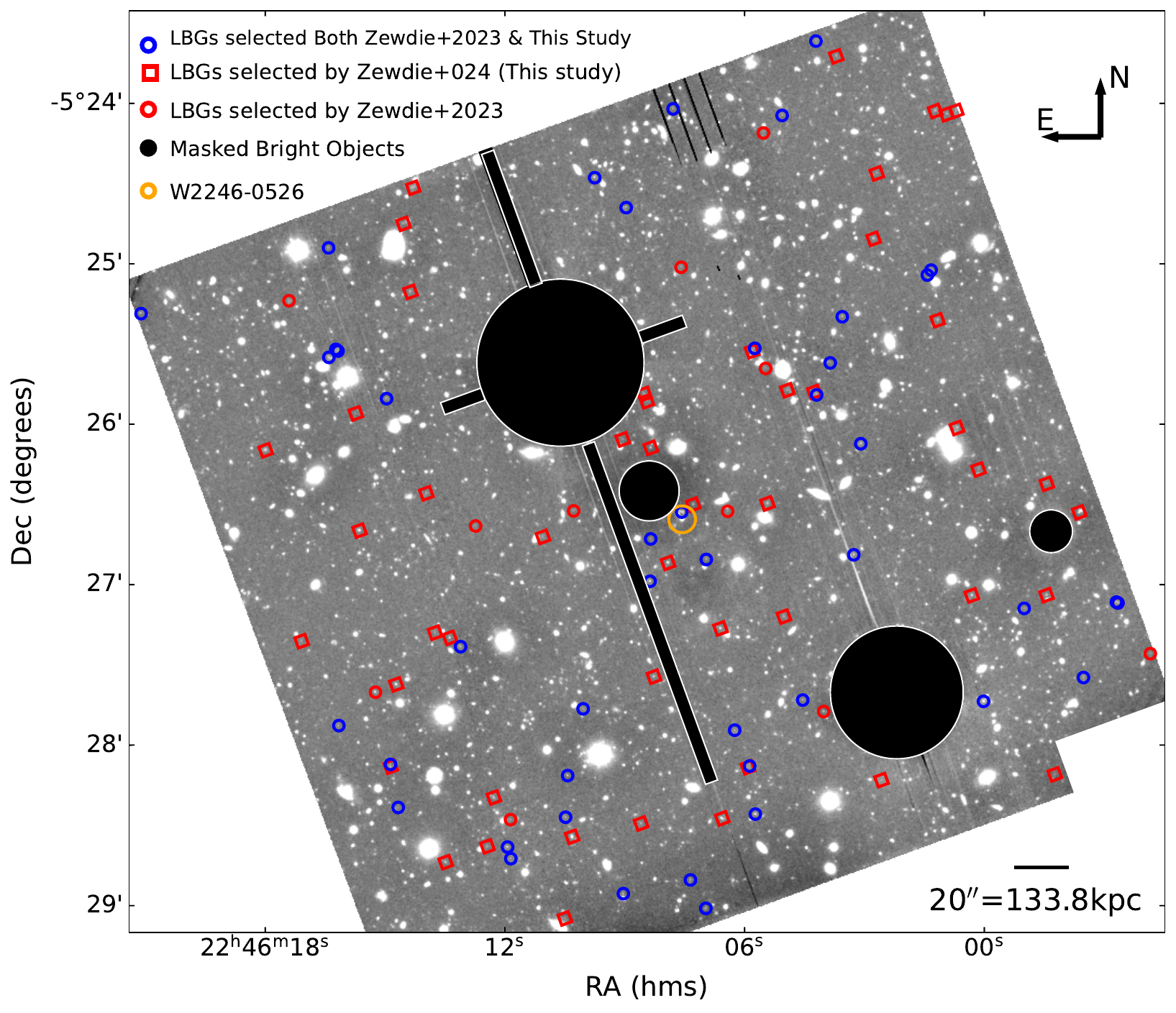} 
\caption{Adapted from \citetalias{2023Zewdie}, the \textit{i}-band image of W2246--0526 and the solid orange circle indicate its position. The blue circles show the LBG candidates selected by \citetalias{2023Zewdie} using the modified and with optimised selection criteria.The black circles and rectangles show the masked area. The red circles and rectangles show the LBG candidates only selected by \citetalias{2023Zewdie} and using optimise selection criteria, respectively.}
\label{ds9_w2246}
\end{figure}
 

There are a number of saturated stars, particularly in the field of W0410--0913, with large saturation spikes that cause irregular systematic features in the image, so we applied some conservative masking to avoid the detection of spurious sources near the saturated spikes. The final masked images are shown in Figure \ref{ds9_image}.  We estimated the usable area by generating 10$^6$ random points distributed throughout the image and counted the fraction of unmasked points. The usable area remaining after masking is $\sim$~212.6 arcmin$^2$ and $\sim$235.7  arcmin$^2$ for the W0410--0913 and W0831+0140 fields, respectively. The IMACS instrument is composed of 8 CCDs with a gap between them at the center of the field-of-view. Therefore, we center the targets on one of the CCDs.  We considered the region around the Hot DOG as the center up to a maximum radius. We refer to this area as the ``inner region'' as the which has a radius of $\sim$4.8\arcmin.


\subsection{Photometry}\label{Photometry}

We measured the photometry using fixed 2\arcsec diameter apertures with {\tt{SExtractor}}\footnote{{\tt{SExtractor}}  version 2.28 } \citep{1996Bertin}  in dual-image mode, using the \textit{r}-band images for source detection. We used detection and analysis thresholds of 3 pixels detected above 1.5$\sigma$. We applied a 5$\times$5 convolution filter based on a Gaussian point spread function (PSF) with a full width at half maximum (FWHM) of 3.0 pix (0.66\arcsec).  For the background, we used a global model with mesh and filter sizes of 32 and 3 in pixels, respectively.


Photometric calibration was conducted using data from the Panoramic Survey Telescope \& Rapid Response System (PanSTARRS) Survey \citep{2012Tonry}. We exclusively considered point sources, which were selected based on the probabilistic classification of unresolved point sources with a {\tt{ps\_score}} greater than 0.83, following the suggestion by \cite{2018Tachibana}\footnote{ \url{https://outerspace.stsci.edu/display/PANSTARRS/How+to+separate+stars+and+galaxies}}. The PanSTARRS point sources were cross-matched with our sources using a 1\arcsec\, radius, resulting in 175 and 372 matches within the unmasked areas of the W0410--0913 and W0831+0140 fields, respectively. To address potential issues with saturation and non-linearity in the IMACS images, as well as issues with low signal-to-noise ratio (SNR) detections in PanSTARRS, we only considered PanSTARRS point sources within the PSF magnitude ranges of $18.0 < g < 22.5$,  $17.5 < r < 21.5$,  and $17.0 < i < 21.0$, resulting in 71 (201), 78 (205) and 77 (204) sources, respectively, for W0410--0913 (W0831+0140). To calibrate the IMACS \textit{g}-band observations, we found that a single colour term, $g-r$ from PanSTARRS, was needed in addition to the PanSTARRS \textit{g}-band magnitude. For the IMACS \textit{r}-band, we found that two colour terms were needed, so we used the $g-r$ and $r-i$ PanSTARRS colours. Similarly, for the IMACS \textit{i}-band, we found that we needed to use the $r-i$ and $i-z$ PanSTARRS colours. The 1, 3, and 5$\sigma$ depths of the image stacks are presented in Table \ref{observation}.

We also used deep GMOS imaging data in the \textit{r}-, \textit{i}-, and \textit{z}-bands of Gemini GMOS-S presented by \citetalias{2023Zewdie} in the field around the W2246--0526. The magnitude limits of the stacked images for the 5$\sigma$, 3$\sigma$, and 1$\sigma$ depths of the \textit{i}- and \textit{z}-bands are detailed in Table \ref{observation}. Using a similar criteria for masking as was done for the IMACS observations resulted in a final FoV of 23.7 arcmin$^2$ as shown in Figure \ref{ds9_w2246}. For further details, we refer the reader to \citetalias{2023Zewdie}.

\section{LBG selection function optimisation}\label{optimazation}

\subsection{COSMOS data} \label{COSMOS_data}

 \begin{figure}
\centering
\includegraphics[scale=0.39]{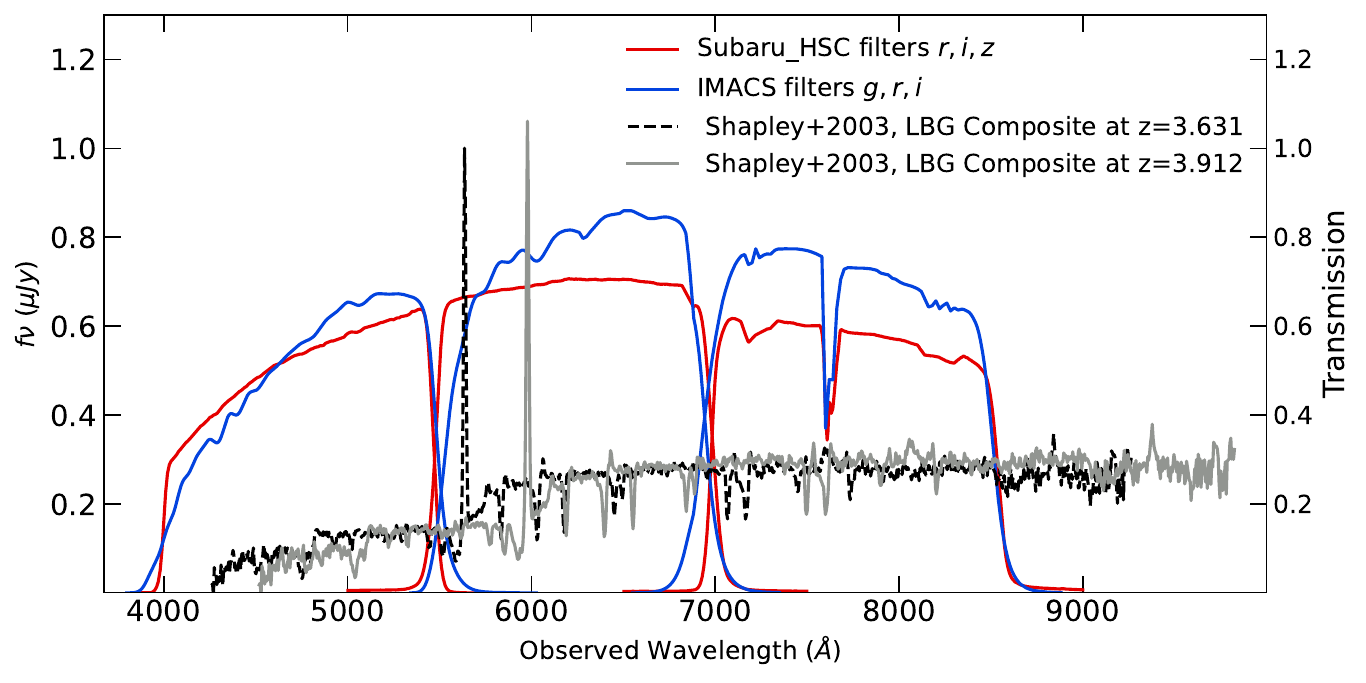}
\caption{The composite LBG spectrum from \cite{2003Shapley} shifted to $z = 3.631$ (black dashed lines) and $z = 3.912$ (gray solid lines), used for optimising the selection function for the W0410--0913 and W0831+0140 fields, respectively. After accounting for quantum efficiency and atmospheric transmission, the red and blue lines represent the HSC filter curves (used for optimisation and the blank field) and the IMACS filter curves (used for W0410--0913 and W0831+0140), respectively.}
\label{spectra}
\end{figure}

The COSMOS2020 field offers a unique multiwavelength dataset covering a relatively large area of $\sim$2 deg$^{2}$. Here, we use the COSMOS2020 CLASSIC catalog \citep{2022Weaver}\footnote{The COSMOS2020 catalog is available for download at \url{https://cosmos2020.calet.org/}}, where source detection was conducted using {\tt{SExtractor}}. The catalog provides observations in multiple broad and medium photometric bands. Additionally, it provides spectroscopic and photometric redshift estimates obtained with {\tt{LePhare}} \citep{2002Arnouts, 2006Ilbert} and {\tt{EAZY}} \citep{2008Brammer} for more than 1.7 million sources. \cite{2022Weaver} quantify the uncertainty in photometric redshift estimates using the normalized median absolute deviation \cite[NMAD,][]{1983Hoaglin},  defined as  $\rm \sigma_{NMAD} = 1.48 \times median \left(\frac{|\Delta z - median(\Delta z)|}{1+z_{spec}}\right)$. In COSMOS2020, the precision of photometric redshifts (expressed as $\rm \sigma_{NMAD}$) is typically subpercent, about $0.01(1 + z)$ for sources with $i < 21$. For the faintest sources ($25 < i < 27$), the precision remains around 5\%, and it stays better than $0.025(1 + z)$ for sources with $i < 25$.

In this work, we use the photometry and photometric redshift estimates in the COSMOS2020 catalog to optimise the LBG selection function for each of the Hot DOG fields studied, as discussed in the next section. We specifically use the Subaru HSC photometry in the \textit{g}-, \textit{r}-, \textit{i}- and \textit{z}-bands which have  3$\sigma$ depths of 28.1, 27.8, 27.6, and 27.2 mag, respectively, within 2\arcsec\, apertures. We only considered sources whose photometry is not affected by saturated stars and their spikes. We further require sources to be in the overlap region of the UltraVISTA, HSC and SuprimeCam imaging (i.e., FLAG\_COMBINED, clean=0;  hereafter, combined catalog, see \citealt{2022Weaver}) to ensure a uniform quality of photometric redshifts and object classifications. The area of this combined region is 1.278 $\rm deg^2$.  The combined COSMOS2020 catalog has a total of 723,897 sources and includes spectroscopic or accurate photometric redshift estimates, as well as  separate classifications for galaxies (711,918), stars (9,644), X-rays sources (2,170), and failure sources (165), which enable us to estimate the contamination level. As shown in Figure \ref{spectra}, the filter curves are very similar between the instruments. Hence, it is not required to calculate the colour differences between them.

\begin{figure*}
\includegraphics[scale=0.6]{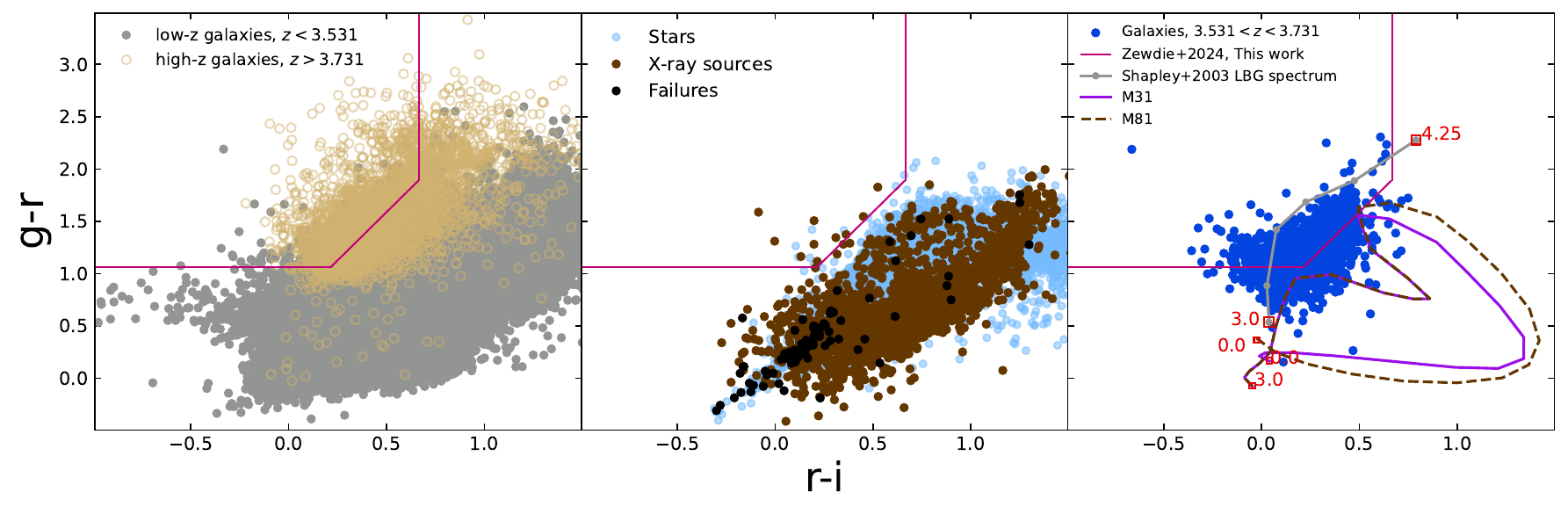}
\caption{$g-r$ vs $r-i$ colour distribution of simulated magnitude sources from the combined catalog in the COSMOS2020 field used to optimise the selection function of companions to W0410--0913 in the redshift range $3.531<z<3.731$. The left panel represents galaxies at lower redshifts ($z<3.531$, gray) and higher redshifts ($z>3.731$, tan). The middle panel displays stars, X-ray sources, and sources with failed photometric redshift measurements, while the right panel shows galaxies in the targeted redshift range of $3.631 \pm 0.1$. The solid magenta line represents the optimised selection function at $3.531<z<3.731$. The gray line shows the colour-redshift track of the LBG composite spectrum of \protect\cite{2003Shapley} with the IGM absorption of \protect\cite{1995Madau} shifted from $z=3.0$ to $z=4.25$, where the dots indicate $\Delta z= 0.25$. We show the representative colours of several classes of galaxies from \protect\cite{1980Coleman} as a function of redshift from $z= 0$ to $z= 3.0$.  }
\label{W0410_FLAG_color_simulated}
\end{figure*}

\begin{figure*}
\includegraphics[scale=0.6]{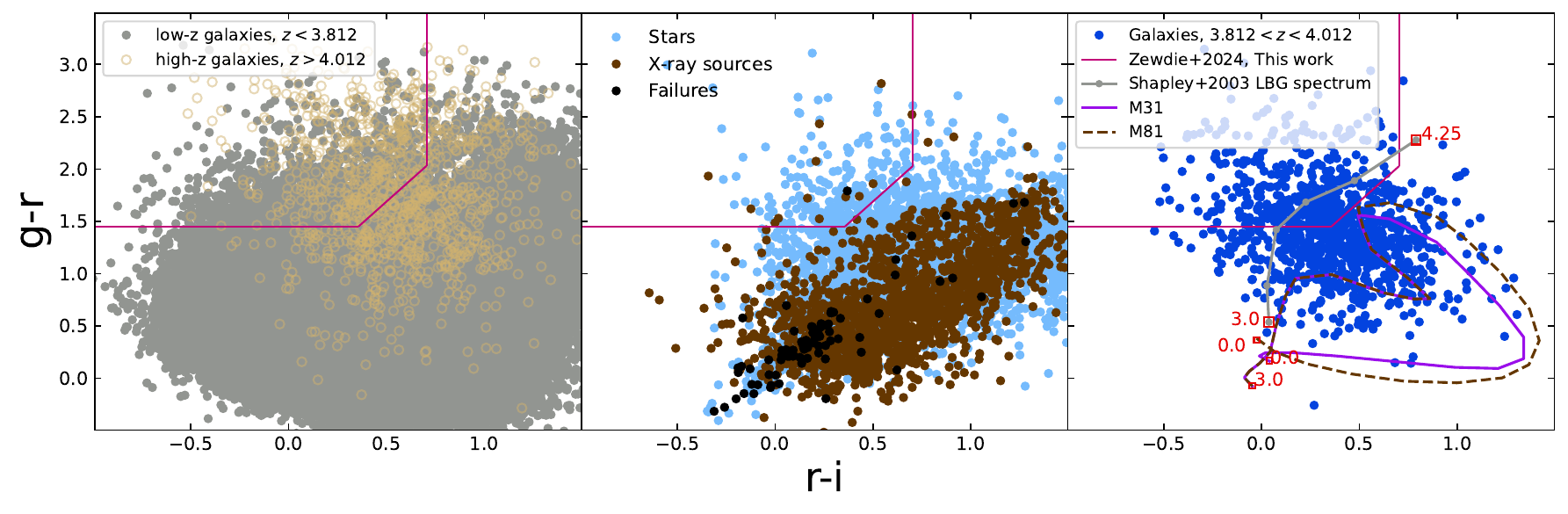}
\caption{Same as Figure \ref{W0410_FLAG_color_simulated} but for optimising the selection of companions to W0831+0140 in the redshift range $3.812<z<4.012$.}
\label{W0831_FLAG_color_simulated}
\end{figure*}

\begin{figure*}
\includegraphics[scale=0.6]{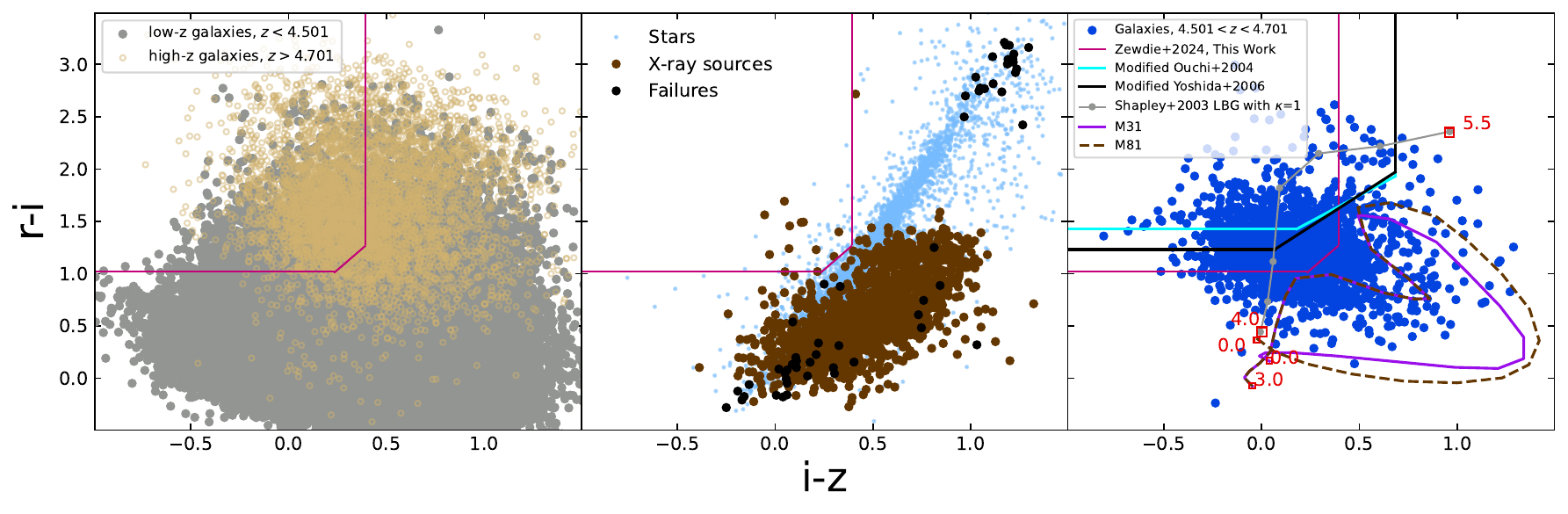}
\caption{$r-i$ vs $i-z$ colour distribution simulated magnitude sources from the combined catalog in the COSMOS2020 field used to optimise the selection of companions to W2246--0526 in the redshift range $4.601\pm 0.1$. Symbols are defined in the same way, as described in detail in Figure \ref{W0410_FLAG_color_simulated}. The figure also shows the selection function adopted by  \protect\citetalias{2023Zewdie} based on those of \protect\citet[cyan]{2004VOuchi} and the \protect\citet[black; See \protect\citetalias{2023Zewdie} for details]{2006Yoshida}. The gray line is the colour-redshift track of the LBG composite spectrum of \protect\cite{2003Shapley} with the IGM absorption of \protect\cite{1995Madau} shifted from $z=4.0$ to $z=5.5$, the symbols are the same as in Figure \ref{W0410_FLAG_color_simulated}.}
\label{W2246_FLAG_color}  
\end{figure*}

\subsection{LBG selection function}
LBGs are actively star-forming galaxies and, as such, have intrinsically blue spectral energy distributions (SEDs) in the rest-frame UV down to the wavelength of the Ly$\alpha$ emission line. Short of that, their SEDs are significantly depressed by intergalactic Ly$\alpha$ absorption and by the Lyman break shortward of 912 \AA. Photometric identification of LBGs is typically done using three photometric bands. The color between the two bluest bands is used to identify the drop in flux due to the Lyman break, while the color between the two redder bands maps the rest-frame UV continuum of the galaxy.


For the two lower redshift targets, W0410--0913 and W0831+0140, we identify companion galaxies using $g-r$ and $r-i$ colours, while for W2246--0526 we use $r-i$ and $i-z$ colours instead. The exact colour limits one uses determines the purity and the completeness of the LBG sample. In order to optimise the selection function, we used the redshift estimates and the HSC \textit{g}, \textit{r}, \textit{i}, and \textit{z} photometry  of COSMOS2020 sources in the combined COSMOS2020 catalog (see Section \ref{COSMOS_data}). The HSC observations of the COSMOS2020 field are deeper than our IMACS  and GMOS-S observations. Specifically, the depth of COSMOS2020 surpasses our IMACS depths by 1.0, 1.5, 2.0 mag in the \textit{g}-, \textit{r}-, and \textit{i}-bands. In contrast, compared to GMOS, the COSMOS2020 catalog is $\sim 0.5$ magnitude deeper in  the \textit{i}- and \textit{z}-band, while the depth in the \textit{r}-band is comparable. To account for the different depths of the COSMOS2020 catalog and our datasets, we added noise to the HSC COSMOS2020 photometry to match that of our observations. We modeled the photometric uncertainty as a function of magnitude in our IMACS and GMOS-S field as:

\begin{equation}\label{mag_fit}
    \delta m (m) =Ae^{\beta m}
\end{equation}
where $\delta$m is the magnitude error, $m$ is magnitude, and $A$ and $\beta$ are constants we fit for.

Table \ref{constants_bA} shows the best-fit $\beta$ and $A$ values for each filter in each Hot DOG field. We note that at the background-dominated limit, one would expect $\beta=0.92$, which is very close to the best-fit values. Using this relation, we create a simulated version of the COSMOS2020 data, matching the depth of each band of the IMACS and GMOS-S fields. Specifically, for every object in COSMOS2020, we simulate a new magnitude for COSMOS2020 sources brighter than the 1$\sigma$ depth of the field in question. The new magnitudes are randomly drawn from a Gaussian distribution with a mean equal to the COSMOS2020 HSC magnitude in the respective band and a dispersion equal to ($\delta m^{2}(m) - \delta m^2_{\rm HSC})^{1/2}$, where $\delta m_{\rm HSC}$ is the photometric uncertainty of the HSC observations.  

\begin{table}
    \centering
    \caption{Constants to model the photometric uncertainty as a function of magnitude in our fields.}
    \label{constants_bA}
    \begin{tabular}{c c c c}
        \hline
        Hot DOGs & filters & $\beta$ & $A (10^{-11}$)\\
                \hline
         & \textit{g} & 0.903 & 0.789\\
 W0410--0913       & \textit{r} & 0.896 & 2.031\\
        & \textit{i} & 0.901 &3.396 \\
        \hline
         & \textit{g} & 0.906 &0.940 \\
 W0831+0140       & \textit{r} & 0.891 & 3.227\\
        & \textit{i} & 0.902 & 3.386\\
                \hline
         & \textit{r} & 0.919 &0.274 \\
 W2246--0526       & \textit{i} & 0.920 & 0.519\\
        & \textit{z} & 0.920 & 0.850\\
    \end{tabular}
\end{table}

We then proceed to optimise the LBG selection function separately for each of our fields using this modified COSMOS2020 photometry.  The optimisation of the photometric selection criteria only considere sources fainter than the Hot DOG in each field in the reddest band used for the colour selection (i.e., $i=23.96$ for W0410--0913, $i=22.32$ for W0831+0140, and $z=22.31$ for W2246--0526). Since LBG are unlikely to be brighter than the Hot DOG, it helps ensure the robustness of the LBG selection and minimises contamination. Additionally, we only use sources with magnitudes brighter than the 3$\sigma$ depth of our images in the \textit{r(i)}- and \textit{i(z)}-bands, and brighter than the 1$\sigma$ depth of our images in the \textit{g(r)}-bands in the IMACS (GMOS) observations.

We assume a general shape of the selection functions based on those of \cite{2004VOuchi}. Specifically, we require that: i) sources are red in the bluest colour ($g-r$ or $r-i$ depending on the field) to target the depression in the SED caused by the Lyman break and the intergalactic medium (IGM) absorption; ii) sources are blue in the reddest colour ($r-i$ or $i-z$ depending on the field) to ensure the continuum redwards to Ly$\alpha$ is consistent with a star-forming SED; and iii) that they meet a joint colour threshold to avoid contamination from lower redshift galaxies. We optimise the selection function by maximizing the contrast of the number of galaxies in the intended redshift range ($\rm N_{\rm Targ}$) with respect to contaminants. Specifically, we select colours that maximize the function:

\begin{equation} \label{eqn_con}
\Theta =\rm \frac{N_{\rm Targ}}{(N_{\rm Targ} + N_{{\rm low} z} + N_{{\rm high} z} + N_{\rm Stars} + N_{\rm X-ray} + N_{\rm Fail})^{1/2}}, 
\end{equation}

\noindent where $\rm N_{{\rm Targ}}$ are the galaxies in the targeted redshift range, defined as the redshift of the Hot DOG +/- 0.1, $\rm N_{{\rm low} z}$ and $\rm N_{{high} z}$ are the galaxies with redshifts below and above the targeted range, respectively,  $\rm N_{{\rm Stars}}$ are stars, $\rm N_{{\rm X-ray}}$ are X-ray sources, and $\rm N_{{\rm Fail}}$ are the failures, for which the photometric redshift fit  failed \cite[most of these objects have photometry from only a single band, see][]{2022Weaver}.

We find that the optimal selection function for sources at the redshift of W0410--0913 ($z=3.163$) is given by:

\begin{align}\label{eqW041}
 	g-r & >1.064,  \nonumber\\
	r-i & <0.669,  \nonumber \\
 	g-r & >1.868(r-i)+0.65.
\end{align}\\



\noindent For sources at the redshift of W0831+0140 ($z=3.912$) it is given by:
\begin{align}\label{eqW081}
 	g-r & >1.45,  \nonumber\\
	r-i & <0.705,  \nonumber \\
 	g-r & >1.668(r-i)+0.855. 
\end{align}\\

\noindent And for W2246--0526 ($z=4.601$) we find:
\begin{align}\label{eqW221}
  r-i & >1.017,  \nonumber\\
  i-z & <0.393,  \nonumber \\
  r-i & >1.611(i-z)+0.634. 
\end{align}
 
Figures \ref{W0410_FLAG_color_simulated}, \ref{W0831_FLAG_color_simulated} and \ref{W2246_FLAG_color} show the colour distribution of COSMOS2020 sources with the modified magnitudes used to optimise the LBG selection functions for the redshifts of W0410--0913, W0831+0140, and W2246--0526, respectively. The right panels of the figures also show the colours of representative galaxy templates from \cite{1980Coleman} in the redshift range of 0-3, and the LBG composite spectrum from \cite{2003Shapley}. We used the \cite{1995Madau} model, assuming the mean IGM optical depth for the Hot DOG redshift.

We estimated the optimal reliability and completeness within the COSMOS fields. We found that the reliability of our optimal selection function for W0410-0913, W0831+0140, and W2246-0526 is 12.9\%, 12.3\%, and 13.3\%, respectively. As can be seen in the left panel of Figures \ref{W0410_FLAG_color_simulated}, \ref{W0831_FLAG_color_simulated}, and \ref{W2246_FLAG_color}, a higher fraction of the contaminants are galaxies within 0.2 and 0.5 units of redshift. We found that the completeness of our optimal selection function for W0410-0913, W0831+0140, and W2246-0526 is 44.6\%, 34.4\%, and 61.8\%, respectively, although we remark the selection is optimized for contrast and not independently for reliability or completeness.

\section{Results and Discussion} \label{sec_LBGc}
\subsection{Lyman-break galaxy candidates}
\begin{figure*}[h!]
\includegraphics[scale=0.5]{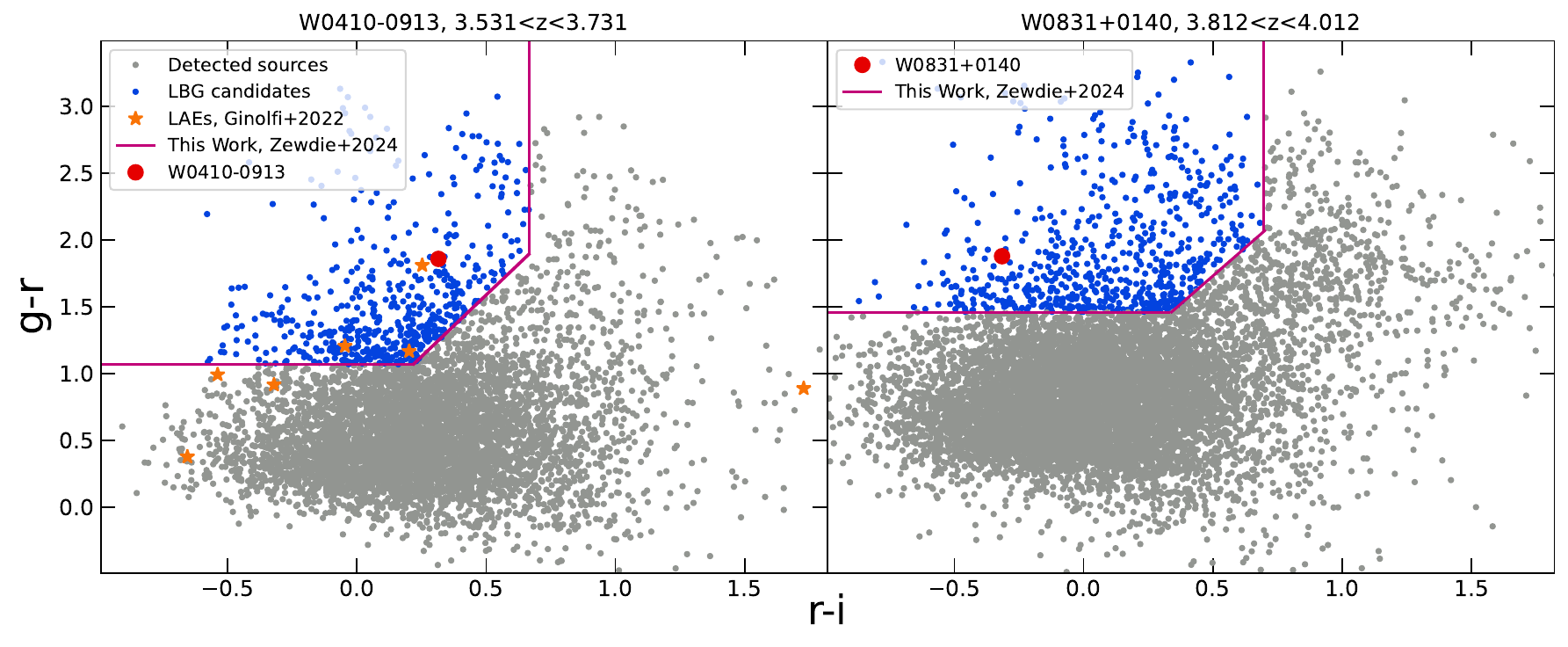} 
\caption{Distribution of $g-r$ vs $r-i$ colours of sources around W0410--0913 at $z = 3.631$ (left panel) and W0831+0140 at $z = 3.912$ (right panel). The Hot DOGs are denoted by red filled-circles.  In both panels, gray dots represent detected sources, and the blue dots represent the LBG candidates. The magenta lines show the optimised selection criteria based on the simulated magnitude sources from combined COSMOS2020 (see Section \ref{optimazation}). In the left panel, the orange filled stars are the LAEs detected by VLT/MUSE observations \citep{2022Ginolfi}.}
\label{W0831_LBG}
\end{figure*}

\begin{figure}
\includegraphics[scale=0.4]{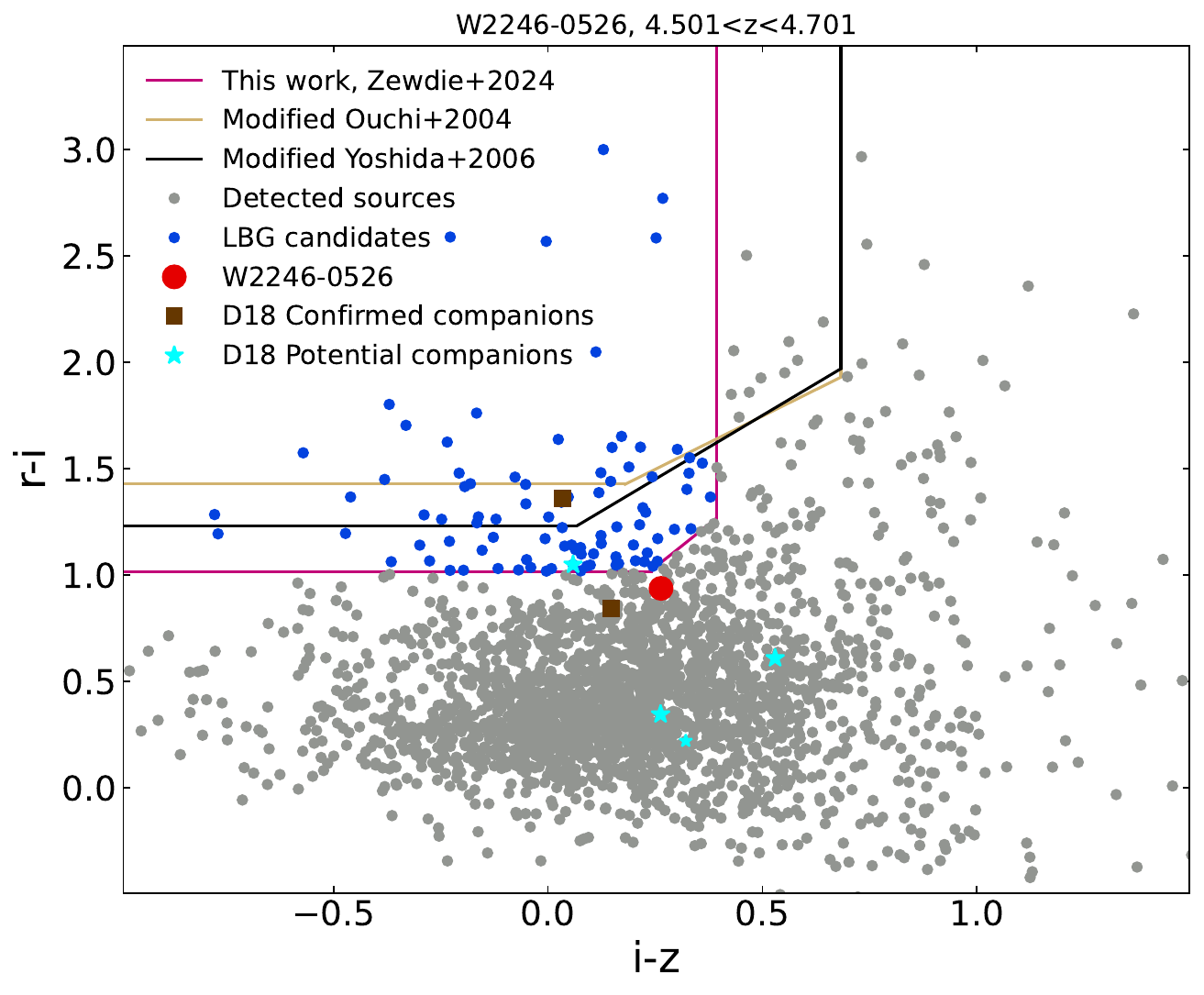}
\caption{Distribution of $r-i$ vs $i-z$ colours of sources around W2246--0526 at $z = 4.601$. The red filled circle, gray and blue dots, and the magenta line mean the same as in Figure \ref{W0831_LBG}. Brown filled squares and cyan filled stars are the confirmed and potential companions detected with ALMA observations \citep{2018Tanio}. The description of the selection function are the same as in Figure \ref{W2246_FLAG_color}.}
\label{W2246_LBG}
\end{figure}





We applied the optimised selection functions described in the previous section to select LBG candidates around each Hot DOG's field. We eliminated sources brighter than the Hot DOGs in the \textit{i}-band and fainter than 3$\sigma$ in the \textit{r} and \textit{i}-bands. Sources fainter than the 1$\sigma$ magnitude limit in the \textit{g}-band are treated as upper limits (1$\sigma$) for W0410--0913 and W0831+0140. Similarly, for W2246--0526, we eliminated sources brighter than the Hot DOG in the \textit{z}-band, as well as those fainter than 3$\sigma$ in the \textit{i} and \textit{z}-bands. Sources fainter than the 1$\sigma$ magnitude limit in the \textit{r}-band are treated as upper limits (1$\sigma$). We found 549, 676, and 96 LBG candidates around W0410--0913, W0831+0140, and W2246--0526, respectively. As mentioned earlier, the Hot DOGs were not positioned at the center of the IMACS FoV. Within the inner region (see Figure \ref{ds9_image} and Section \ref{ODR}), and we found 182 and 184 LBG candidates in the field of W0410--0913 and W0831+0140, respectively. These numbers are summarized in Table \ref{Table_LBG_sim}.

Figures \ref{W0831_LBG} and \ref{W2246_LBG} show the colour distributions of detected objects in each field, highlighting the LBG candidates. Figure \ref{W2246_LBG} also shows the selection functions used by \citetalias{2023Zewdie} to identify LBG candidates around W2246--0526, adapted from the studies of \cite{2004VOuchi} and \cite{2006Yoshida}. The former identified 37 LBG candidates, while the latter identified 55. The optimised selection function determined in the current work identifies 96 LBG candidates over the same area. A direct comparison is difficult, however, as the different selection functions are likely affected by different levels of completeness and reliability. We note, however, that W2246--0526 is not selected as an LBG in our study, nor by either selection function considered by \citetalias{2023Zewdie}.  This outcome likely arises from its unique SED which is influenced by strong dust and AGN activity. However, the other two Hot DOGs are selected as LBG candidates (Figure \ref{W0831_LBG}). Unlike W2246--0526, both W0410--0913 and W0831+0140 are selected by our criteria as LBGs although we note the former is close to the edge of our optimised selection region (see Figure \ref{W0831_LBG}).



\subsection{Overdensity of LBGs around the Hot DOGs}\label{overdensity}

\begin{figure*}
	\includegraphics[scale=0.42]{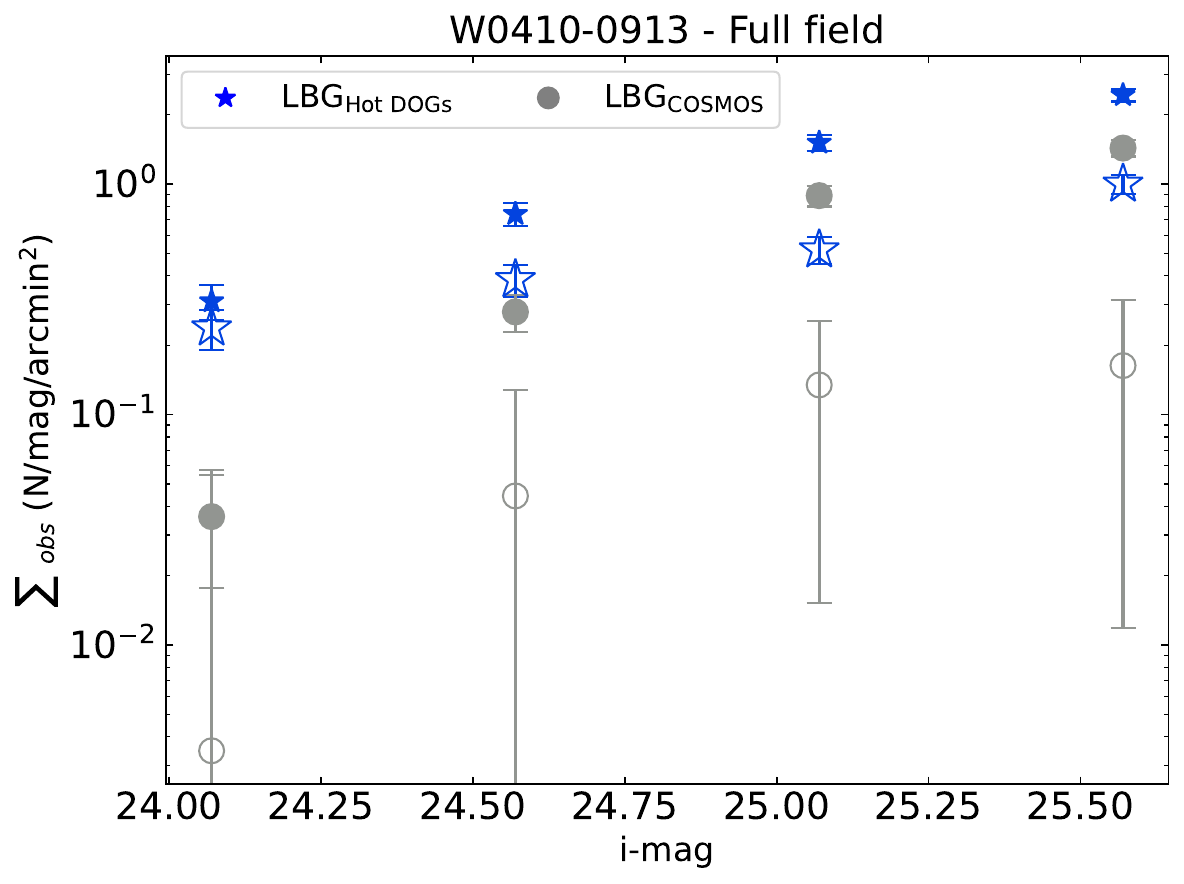}\includegraphics[scale=0.42]{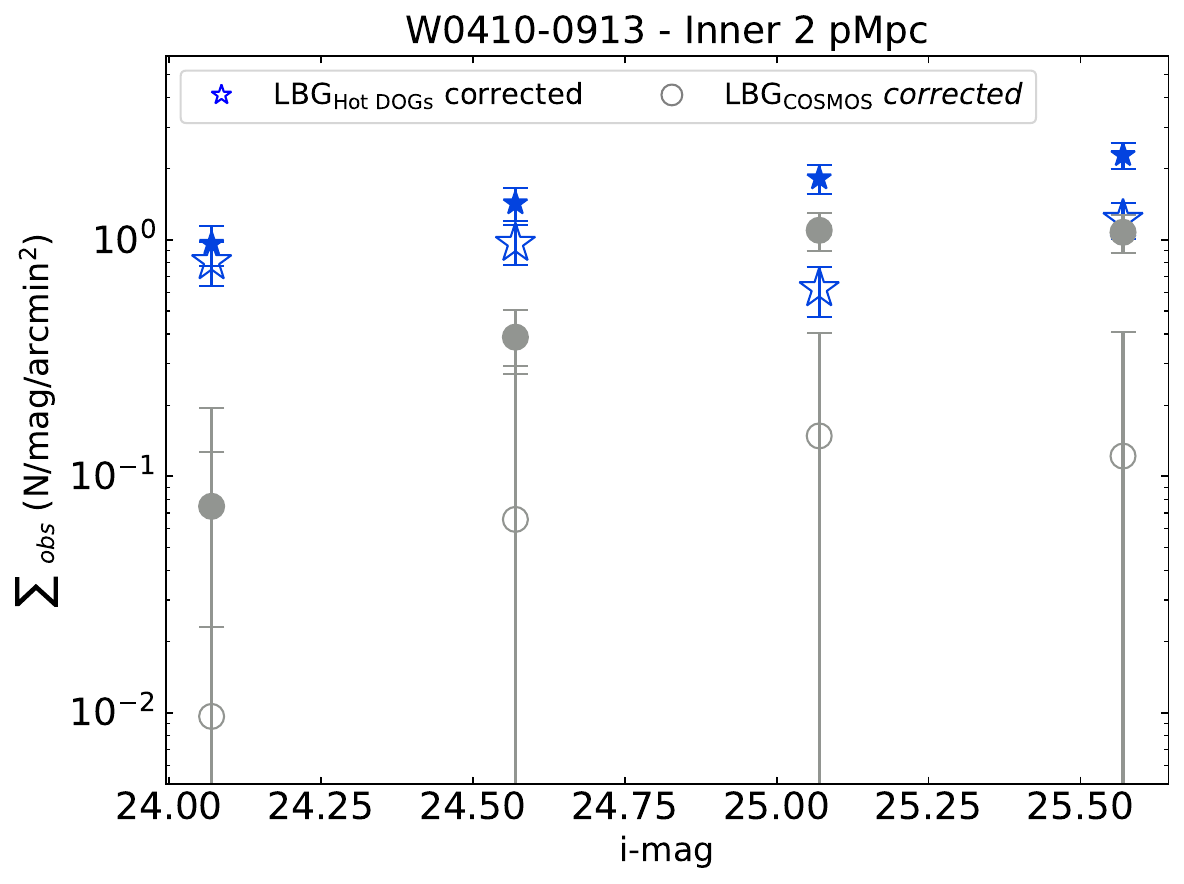}
	\caption{Surface density of LBG candidates around W0410–0913 in the full field (left panel) and in the inner region within the 4.8\arcmin\, the Hot DOGs (right panel). Solid and open symbols show the densities without and with correcting for contaminants, respectively.}
	\label{surface_density_W0410}
\end{figure*}
  
\begin{figure*}
	\includegraphics[scale=0.42]{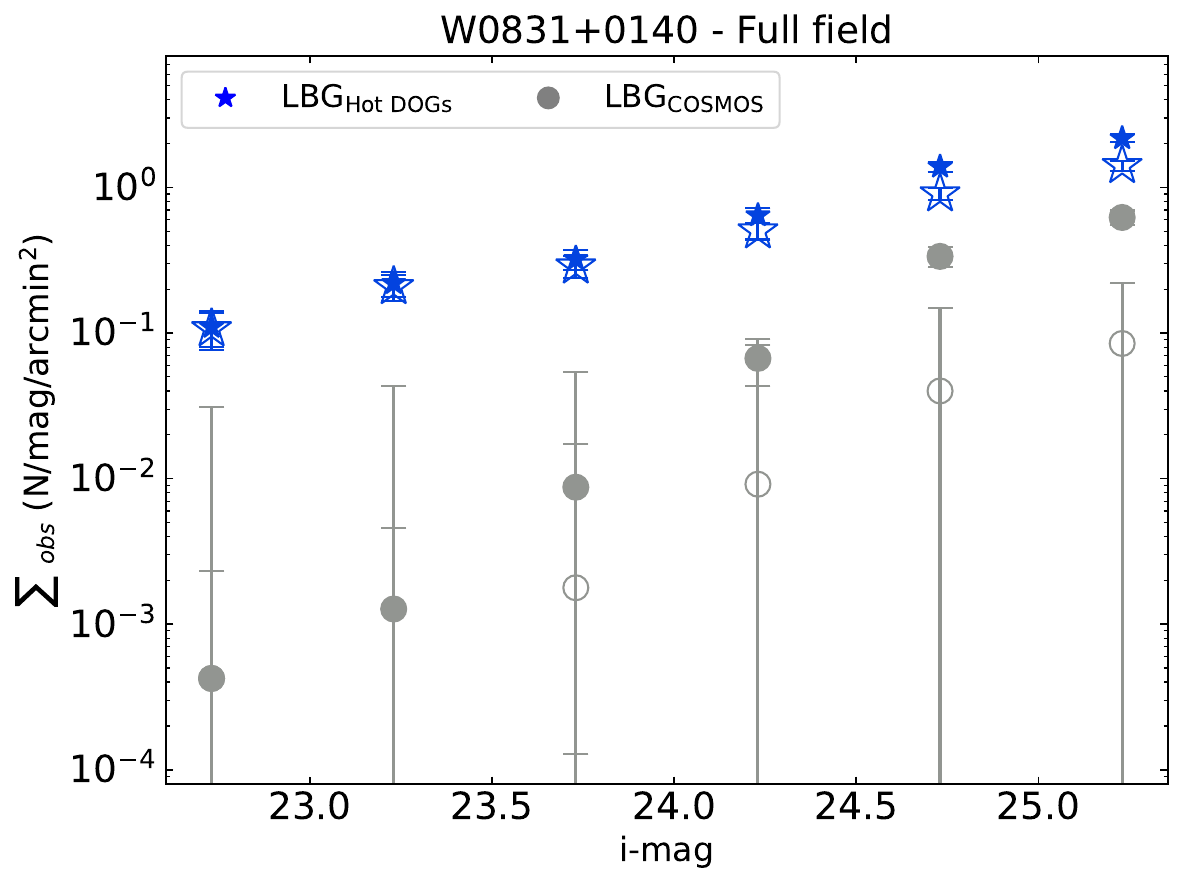} \includegraphics[scale=0.42]{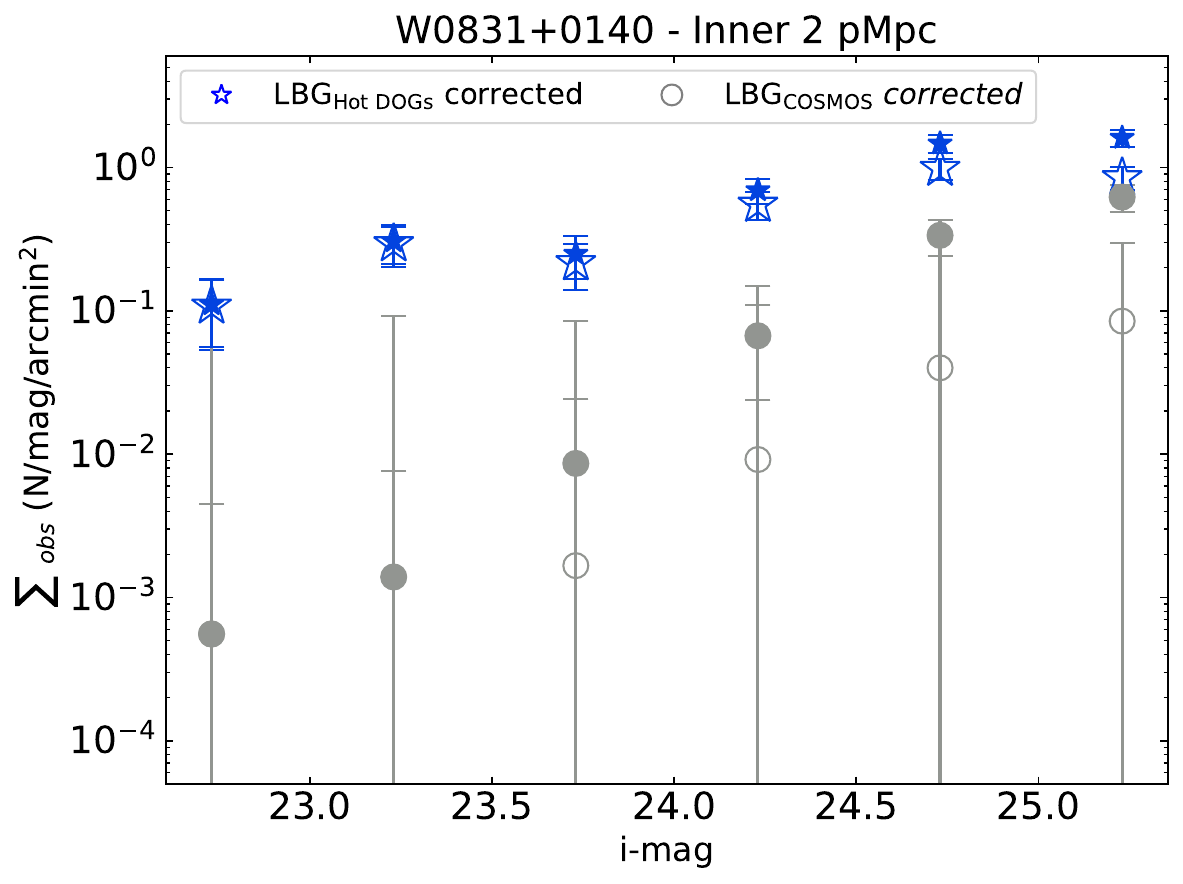}
	\caption{Surface density of LBG candidates around W0831+0140 in the full field (left panel) and in the inner region within the 4.8\arcmin\, the Hot DOG (right panel). Symbools are as in Figure \ref{surface_density_W0410}.}
	\label{surface_density_W0831}
\end{figure*}

\begin{figure}
	\includegraphics[scale=0.42]{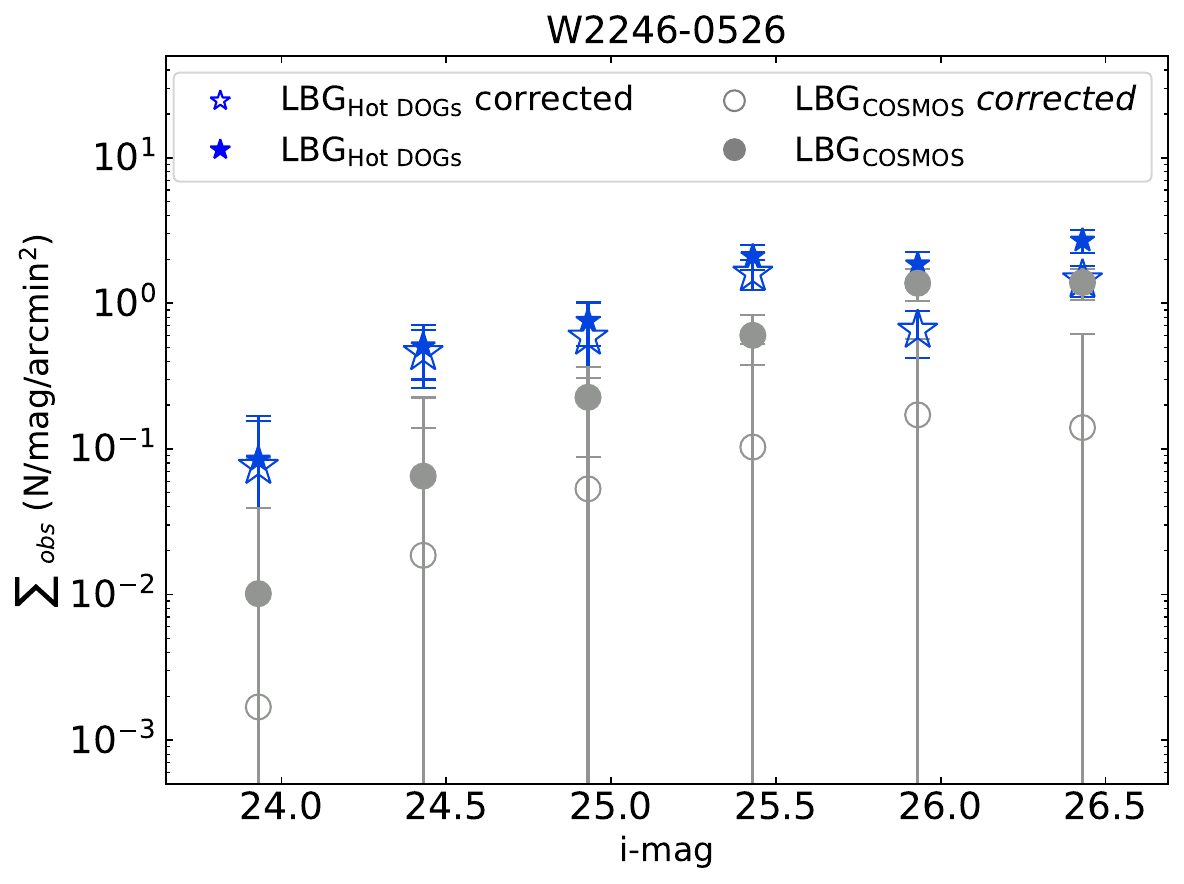}
	\caption{Surface density of the LBG candidates around W2246– 0526. For plot details, see Figure \ref{surface_density_W0410}.}
	\label{surface_density_W2246}
\end{figure}

\begin{table*}
\centering
\caption[]{Statistical information of the environments around the three Hot DOGs studied in this work. We estimate the overdensity using $\delta=\frac{N_F}{N_E}$  and  $\delta^{'} = \rm \frac{N_F - N^C_E}{N_E^{LBGs}}$, and provide the two area values for W0410--0913 and W0831+0140 for the full and inner region. }
\label{Table_LBG_sim}
\begin{tabular}{c c c c c c c }
\hline
\text{\rm Hot DOGs } & \rm Area ($\rm arcmin^2$) & \text{ COSMOS2020 field} & Selected objects &    $\delta$ &  $\delta^{'} $ \\
&  & Selected sources   (Contaminants) & in our field &  &\\
\hline
\noalign{\smallskip}
W0410$-$0913 & ~212.6 & 288 (251) & 528 &  1.83 $\rm \pm 0.08$ & 7.49 $\rm \pm 0.68$ \\
& 56 & 67 (59) & 158 & 2.36 $\rm \pm 0.19$  & 12.38 $\rm \pm 1.65$  \\
\hline
\noalign{\smallskip}
W0831+0140 & 235.7 & 138 (120)      & 645& 4.67 $\rm \pm 0.21$ & 29.17 $\rm \pm 2.21$ \\
& 72 &  42 (37)   & 184 & 4.38 $\rm \pm 0.33$  &  29.40 $\rm \pm 3.84$  \\
\hline
\noalign{\smallskip}
W2246$-$0526 & 23.7 & 39 (34) & 92  & 2.36 $\rm \pm 0.25$ & 11.60 $\rm \pm 1.96$ \\
\hline
\end{tabular}
\end{table*}

As is evident from Table \ref{Table_LBG_sim}, we find a significantly larger number of LBG candidates around Hot DOGs than in the COSMOS2020 field. Assuming the COSMOS2020 field is representative of the average field densities (see below for details), we quantify the overdensities by first comparing the full number of candidates found around each Hot DOG ($\rm N_F$), to the number expected in the same area based on COSMOS2020 ($\rm N_E$), namely:

 \begin{equation} \label{overdensity_eqn1}
    \delta = \rm \frac{N_{F}}{N_E}
\end{equation}
where $\rm {N_E}$ is defined as \(\rm N_{COSMOS}\Psi \), with \(\rm N_{COSMOS}\) being the total number of objects selected in COSMOS by the optimized criteria, \(\rm \Psi=\frac{A_{Hot DOG}}{A_{COSMOS}}\) being the ratio between the area searched in the COSMOS2020 catalog ( \(\rm A_{COSMOS}\)) and the area searched around the given Hot DOG $\rm (A_{HotDOG})$.  This estimate of the overdensity is the simplest, but due to the presence of contaminants it is only a lower limit of the true overdensity. Using the SED classifications and the redshift estimates from the {\tt{LePhare}} models presented in COSMOS2020 we can also try to account for contaminants by estimating the overdensity as: 

\begin{equation}\label{overdensity_eqn2}
    \delta^{'} = \rm \frac{N_F - N^C_E}{N_E^{LBGs}}
\end{equation}

\noindent where \(\rm N_E^{LBGs}\) is defined as \(\rm N^{LBGs}_{COSMOS} \Psi\), with \(\rm N^{LBGs}_{COSMOS}\) being the expected number of galaxies within 0.1 units of redshift from the respective Hot DOG (see eqn. [\ref{eqn_con}]), and \(\rm N^C_{E}\) being defined as \(\rm N^C_{COSMOS}\Psi\), which is the number of contaminants selected by the optimized criteria in COSMOS2020 (which corresponds to all other categories in eqn. [\ref{eqn_con}]). While in principle this should provide a better characterization of the overdensities, it is affected by a number of additional sources of systematic uncertainty (primarily the accuracy of photometric redshift in COSMOS2020) as well as being subject to somewhat arbitrary definitions (e.g.; the targeted redshift range). To provide a more complete picture, we present both estimates for all Hot DOG fields. The true overdensity is expected to be between $\delta$ and $\delta^{'}$, with a higher likelihood of being closer to $\delta^{'}$.

 We estimate the uncertainty of the overdensity as:
\begin{equation}
   \rm \sigma_{\delta} =\rm \frac{N_F}{N_E}  \sqrt{\rm \frac{1}{N_F} + \rm \frac{\Psi}{N_E}}
\end{equation}

\noindent while for the contamination corrected estimate we estimate the uncertainty as:
\begin{equation}
    \rm \sigma_{\delta}^{'} =\rm \frac{1}{N_E^{LBGs}}\sqrt{\rm N_F+N^C_E\Psi + \rm (N_F-N^C_E)^2\frac{\Psi }{N_E^{LBGs}}}
\end{equation}

\noindent We report uncertainties based on Poisson statistics, without added systematic uncertainties to try to account for cosmic variance.

For W0410--0913, considering the entire field of our observations, we find an overdensity of $\delta = 1.83\pm 0.08$ and $\delta^{'} = 7.49\pm 0.68$, while for W0831+0140, we find $\delta = 4.67\pm 0.21$ and $\delta^{'} = 29.17\pm 2.21$. The overdensity factors within the inner regions (see Section \ref{IMACS_data} and Figure \ref{ds9_image}), are $\delta = 2.36\pm 0.19$ and $\delta^{'} = 12.38\pm 1.65$ for W0410--0913, and $\delta = 4.38\pm 0.33$ and $\delta^{'} = 29.50\pm 3.84$ for W0831+0140. For W2246--0526 within the much smaller area probed by the GMOS-S imaging, we find $\delta = 2.36\pm 0.25$ and  $\delta ^{'} = 11.60\pm 1.96$. We note that \cite{2016Assef} showed that W0831+0140 can be classified as a Blue Hot DOG, which are objects whose UV/optical SED is dominated by scattered light from the highly obscured central engine \citep[][]{2016Assef, 2020Assef, 2022Assef}. As such, the host may be significantly fainter than the limit adopted above and could possibly be as faint as the host of W0410-0913. If we only consider LBG candidates fainter than the host galaxy of W0410-0913 (i.e., $i>23.96$) in the W0831+0140, we estimate an overdensity of $\delta=4.67\pm 0.19$ and $\delta^{\prime}=24.67\pm 1.87$. The overdensity factors imply that these Hot DOGs live in very dense environments.

\cite{2022Ginolfi} studied an overdensity of LAEs around W0410--0913 using VLT/MUSE, and they identified 24 LAEs associated with this Hot DOG. In our observations, we detect 10 of these LAEs, although only 7 have the necessary significance in \textit{i}-band to make it into our sample. Of those seven, only three were classified as LBG candidates by our optimised selection function. Of the remaining four, two are very close to the edge of the selection region, while the other two are significantly farther away and may potentially be interlopers.

Figures \ref{surface_density_W0410} - \ref{surface_density_W2246} show the surface density of LBG candidates as a function of the continuum band magnitude (i.e., \textit{i}-band for W0410--0913 and W0831+0140, and \textit{z}-band for W2246--0526). The figures also show the distribution of LBG candidates in COSMOS2020 for comparison. A noticeable trend is observed where the overdensity level seems to diminish towards fainter magnitudes. The diminishing overdensity trend towards fainter magnitudes could be due to several factors, particularly the challenges of detecting faint galaxies and the biases inherent in the selection process. Our observations are not as deep as those from COSMOS, so we are missing faint objects. When we subtract the contamination in both our field and COSMOS, the difference in overdensity becomes apparent in the figures. We optimize the selection using very small redshift bins, which might also contribute to missing faint sources. However, increasing the redshift bins can be lead more contamination.

\begin{figure*}
\includegraphics[scale=0.6]{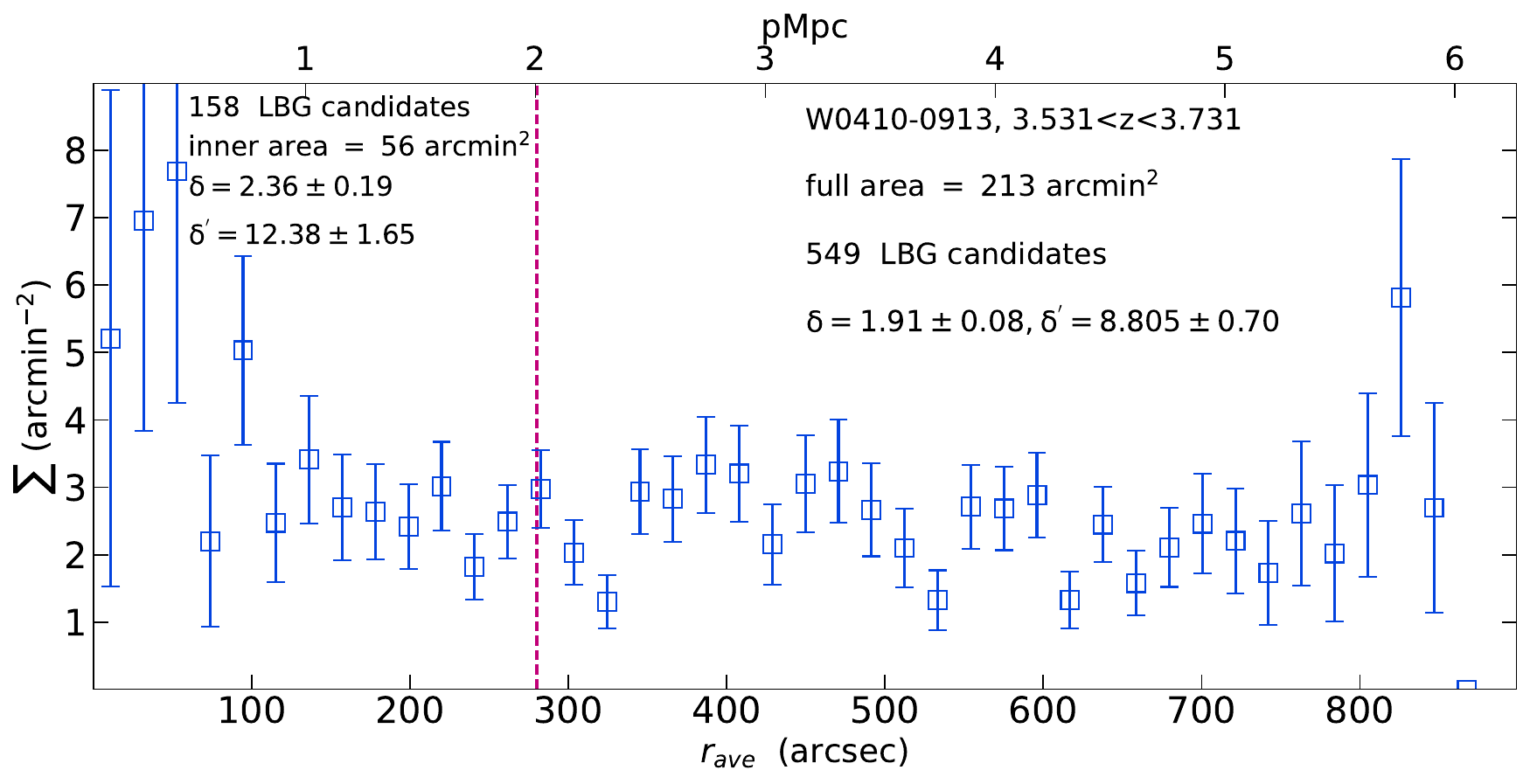}
\caption{Spatial distribution of LBG candidates as a function of distance from W0410--0913. We count the number of LBG candidates in annuli with 20\arcsec radial intervals, avoiding the inner 1\arcsec (7.2 kpc) radius region. The vertical dashed red line represent the 280\arcsec\, radius circle shown in Figure \ref{ds9_image}.}
\label{distance_W0410}
\end{figure*}

  \begin{figure*}
\includegraphics[scale=0.6]{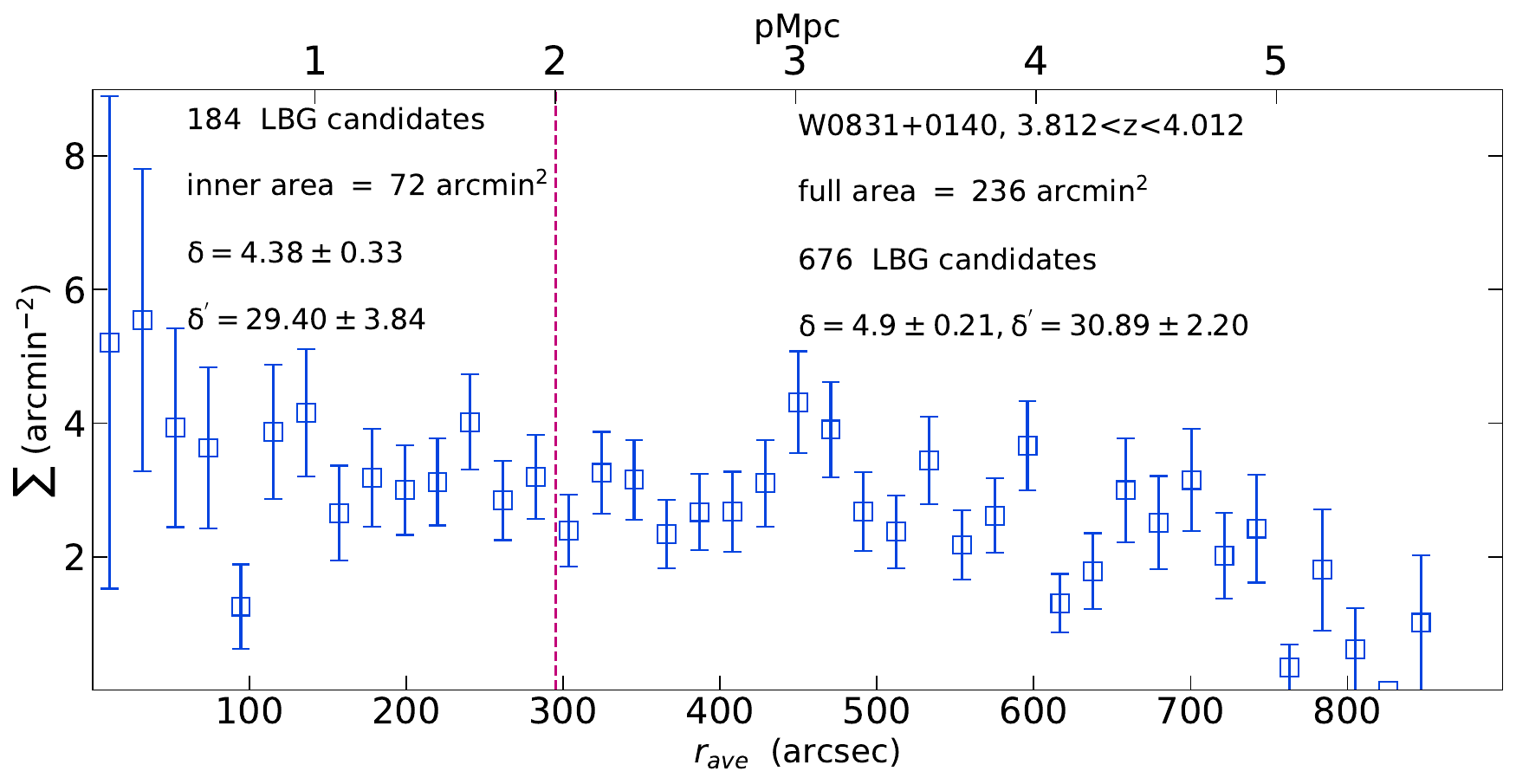}
\caption{Spatial distribution of LBG candidates as a function of distance from W0831+0140. We count the number of LBG candidates in annuli with 20\arcsec radial intervals, avoiding the inner 1\arcsec (7.014 kpc) radius. The vertical dashed red lines represent the 295\arcsec\, radius circle shown in Figure \ref{ds9_image}.}
\label{distance_W0831}
\end{figure*}

\begin{figure*}
\includegraphics[scale=0.6]{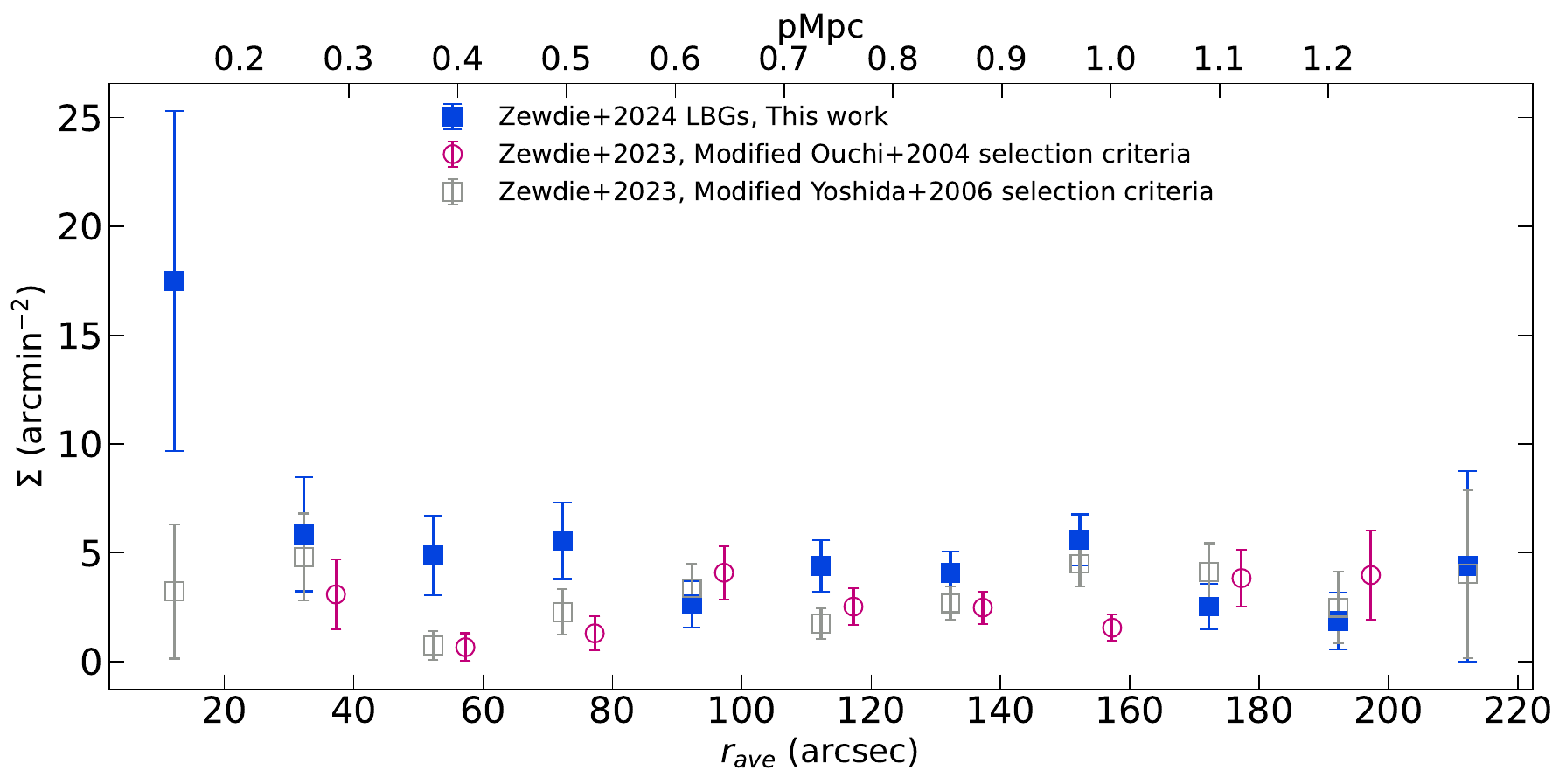}
\caption{Spatial distribution of LBG candidates as a function of their distance from W2246--0526. We count the number of LBG candidates in annuli with 20\arcsec radial intervals, excluding the inner 2\arcsec\, (13.072 kpc) radius. We adapted this analysis from \citetalias{2023Zewdie}. The magenta open circles and gray open squares represent the selected LBG candidates based on modified selection criteria from \citetalias{2023Zewdie}. These selections have been corrected for detection completeness. For clarity, we shifted the surface density of LBG candidates selected by the modified \cite{2004VOuchi} selection criteria by +5\arcsec\, on the x-axis. }
\label{distance_W2246}
\end{figure*}

 \citetalias{2023Zewdie} studied the overdensity of LBGs around W2246--0526 using the Subaru Deep Field (SDF) and Subaru/XMM-Newton Deep Field (SXDF) as blank fields, with slightly modified versions of the selection functions presented by \cite{2004VOuchi} and \cite{2006Yoshida}. These selection functions have negligible levels of contamination (see discussions in the respective articles as well as in \citetalias{2023Zewdie}). They found overdensities of $\delta = 7.1\pm 1.1$ ($\delta = 5.1\pm 1.2$) using the modified selection criteria from \cite{2004VOuchi} and the SDF (SXDF) to determine the expected field densities, and an overdensity of $\delta = 5.2\pm 1.4$ using the modified selection criteria from \cite{2006Yoshida} with the SDF for comparison, resulting in an average overdensity of $5.8^{+2.4}_{-1.9}$. The overdensities found by \citetalias{2023Zewdie} are somewhat lower than what we find in this work, namely $\delta^{\prime} = 11.60 \pm 1.96$. When applying the modified selection of \cite{2004VOuchi} used by \citetalias{2023Zewdie} to the combined region of the COSMOS2020 field, taking into account the magnitude range they used, we find that the COSMOS2020 field has 2.7 (2.5) times higher density of LBG candidates than SDF (SXDF). Similarly, using the modified criteria of \cite{2006Yoshida}, we find COSMOS2020 to have 1.5 times higher density of LBG candidates than SDF. While some of the differences may come from the different filters used (see \citetalias{2023Zewdie} for details), this suggests SDF (which has ~4 times less area  than COSMOS) might be somewhat underdense at the redshift of W2246--0526 ($z=4.601$). We note that all fields involved in this work are far from the Galactic Plane (GP) and Galactic Center (GC), minimizing issues with stellar contamination. Specifically, the fields for W0410--0913, W0831+0140 and W2246--0526 are 53.14, 23.01, and 39.9 deg away from the GP and 74.36, 131.96, and 135.44 deg away from the GC, respectively. For completeness, we note that COSMOS, SDF and SXDF are 42.12, 82.62 and 51.49 deg away from the GP and 113.95, 84.16, and 125.56 deg away from the GC, respectively.

\subsection{Spatial distribution and angular correlation function}\label{LBG_spatial}

Figures \ref{distance_W0410}, \ref{distance_W0831}, and \ref{distance_W2246} show the density of LBG candidates as a function of distance from W0410--0913,  W0831+0140, and W2246--0526, respectively, measured in 20\arcsec\, wide annuli centered on the Hot DOG (not counting the Hot DOG even when selected as an LBG). The overdensity of LBGs shows a profile concentrated around the Hot DOGs, suggesting they correspond to the most massive galaxies in these structures and may become the BCGs of the forming clusters once virialized, as suggested by \cite{2018Tanio}. We note that \citetalias{2023Zewdie} failed to identify a density profile clustering around the Hot DOG (see Figure \ref{distance_W2246}). The difference is likely due to the fact that our optimised selection function has a higher level of completeness and is able to identify many LBGs missed by \citetalias{2023Zewdie} (see the discussion of overdensities around Hot DOGs in their Section 4.1 as well).


Further characterization of the spatial distribution can be achieved by looking at their clustering. We use the two-point angular correlation, $\omega (\theta)$, in each field to provide further evidence that these candidates are truly associated with one another. Specifically, $\omega (\theta)$ is define as the excess probability $\delta P$ of finding objects with an angular separation $\theta$ from each other, such that:
  
\begin{equation}\label{probability}
   \delta P=n^2[1+\omega (\theta)]\delta \Omega_1 \delta \Omega_2
\end{equation}

\noindent where $n$ is the mean number density, and $\delta \Omega_1$  and $\delta \Omega_2$ are elements of solid angle with a separation angle $\theta$.
 
We used the estimator proposed by \cite{1993Landy} to calculate the two-point angular correlation function, namely:

\begin{equation}\label{angular}
   \omega (\theta) = 1+\frac{DD(\theta)}{RR(\theta)}\frac{N_{RR}}{N_{DD}} -2\frac{DR(\theta)}{RR(\theta)}\frac{N_{RR}}{N_{DR}} 
\end{equation}

\noindent where $DD(\theta)$ is the number of pairs of selected LBGs with angular separations between $\theta$ and  $\theta+\Delta\theta$, $RR(\theta)$ is the number of pairs from random catalogs with the same geometry as the selected LBGs, and $DR(\theta)$ is the number of cross-pairs between data and random galaxies. Here, $n_{D}$ and $n_{R}$ are the total number densities of galaxies in the data and random catalogues, while $N_{DD} = n_D * (n_D - 1) / 2$, $N_{RR} = n_R * (n_R - 1) / 2$, and $N_{DR} = n_D * n_R$ are the total numbers of data-data pairs, random-random pairs, and data-random pairs, respectively. This galaxy correlation function estimator is widely used in the literature \citep[e.g.,][]{ 2005Croom, 2006Lee, 2006OverzierC}. We estimate the errors assuming Poisson statistics \citep{2005Croom}. 

\begin{equation}\label{errors}
   \delta\omega(\theta) = \frac{1+\omega(\theta)}{\sqrt{DD_{\theta}}}
\end{equation}

 \begin{table*}
\centering
\caption[]{Summary of the clustering amplitude with the power-law model using best-fitting parameters ($A_{\omega}$ and  $\beta$) and two fixed power-law indices of the correlation function.} 
\label{Table_clustering}
\begin{tabular}{c c c c c c }
\hline
\text{\rm Hot DOGs } & $A_{\omega}^{0.8}$ & $A_{\omega}^{0.6}$ &$A_{\omega}$ &  $\beta$ \\
\hline
\noalign{\smallskip}
W0410$-$0913 & $2.10\pm0.24$ & $0.68 \pm 0.09$  & $13.07\pm3.31$ &  $1.24\pm 0.08$  \\
\hline
\noalign{\smallskip}
W0831+0140 &$ 0.72\pm 0.18$ &$  0.23 \pm 0.07$ & $5.28\pm 2.51$ & $1.30 \pm 0.16$   \\
\hline
\noalign{\smallskip}
W2246--0526 &$ 1.88\pm 0.59$ &$  0.75 \pm 0.26$ & $3.82\pm 3.19$ & $0.99 \pm 0.24$   \\
\hline
\end{tabular}
\end{table*}

\begin{figure}[h!]
    \centering
    \includegraphics[scale=0.45]{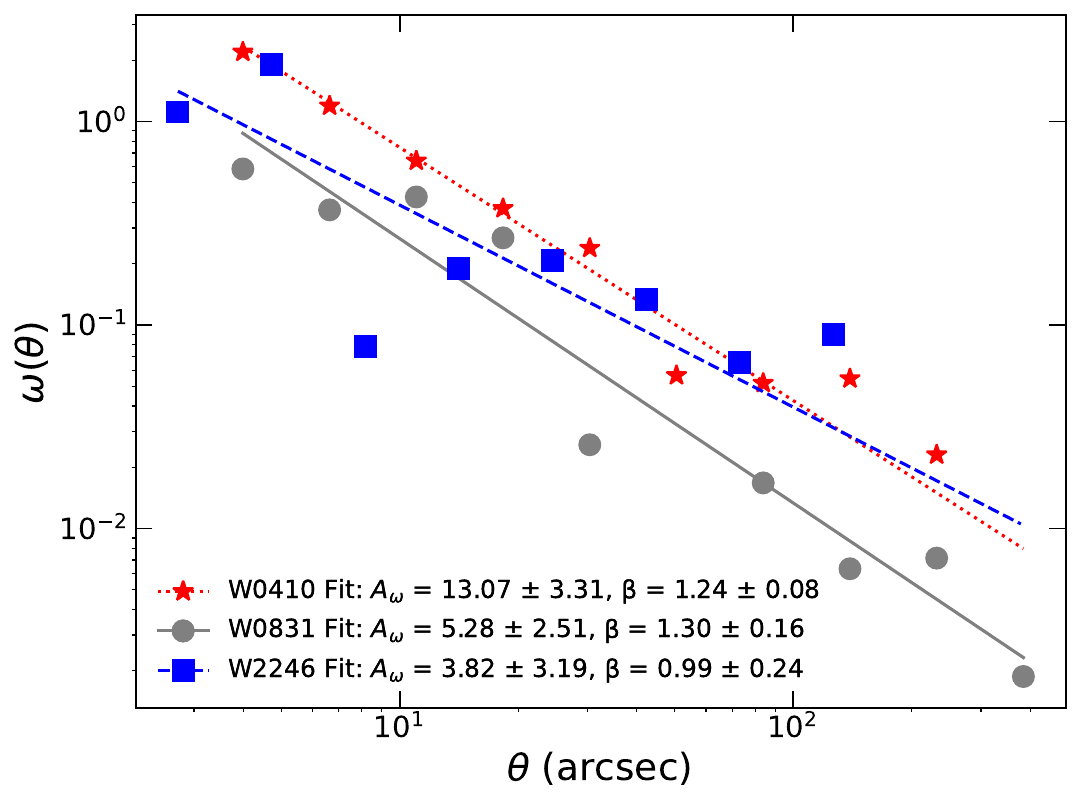}
    \caption{Angular auto-correlation function of the selected LBG candidates around W0410--0913 (red stars),  W0831+0140 (gray solid circles)  and W2246--0526 (blue squares). We used logarithmic binning of separations to ensure sufficient pair counts at small separations. The lines represent the power-law fits (red dotted lines): for W0410--0913 and W0831+0140, the separation angle ranges from 3 arcseconds to 480 arcseconds, with power-law fits (gray solid lines) of \( A_{\omega} = 13.07 \pm 3.31 \) and \( \beta = 1.244 \pm 0.08 \); and  \( A_{\omega} = 5.28 \pm 2.51 \) and \( \beta = 1.30 \pm 0.16 \), respectively. For W2246--0526, the separation angle ranges from 2 arcseconds to 480 arcseconds, with power-law fits (blue dashed lines) of \( A_{\omega} = 3.82 \pm 3.19 \) and \( \beta = 0.99 \pm 0.24 \). }
    \label{clustering_W0410}
\end{figure}

To compute the DR and RR terms we used 10,000 random sources uniformly distributed within an area equivalent that of each field. We then apply the same masks described in Section \ref{ODR} for each field and computed the correlation functions. The results are shown in Figures \ref{clustering_W0410}.

 The angular auto-correlation function is often expressed in a power-law form \citep[e.g., ][]{1999Roche}.

\begin{equation}\label{poweLow}
   \omega(\theta) = A_{\omega}\theta^{-\beta}
\end{equation}

\noindent where $A_{\omega}$ is the amplitude of the auto-correlation function and $\beta$ is slope or power-law index.  As shown in Figures \ref{clustering_W0410}, which show the fit of the angular auto-correlation functions, for W0410-0913, \( A_{\omega} = 13.07 \pm 3.31 \) and \( \beta = 1.244 \pm 0.08 \); for W0831+0140, \( A_{\omega} = 5.28 \pm 2.51 \) and \( \beta = 1.30 \pm 0.16 \) and for W2246--0526, \( A_{\omega} = 3.82 \pm 3.19 \) and \( \beta = 0.99 \pm 0.24 \). We found higher amplitude and slope, indicating strong clustering at smaller scales.

Several analyses have fit the power law by fixing $\beta=0.8$ and $\beta=0.6$. We also fixed the power-law index, $\beta$, value to estimate the clustering amplitude in each field as shown in Table \ref{Table_clustering}. \cite{2001Ouchi} studied the clustering amplitude for three fields at $z \sim 4$, and \cite{2016Harikane} studied three fields at $3.8 < z < 6.8$ by fixing $\beta = 0.8$ and found that $A_{\omega}$ ranged from $0.56 \pm 0.25$ to $0.97 \pm 0.57$ and from $0.2 \pm 0.10$ to $8.8 \pm 3.4$, respectively. Similarly, \cite{2006Lee} studied ten fields at $3.5 < z < 5.5$ by fixing $\beta = 0.6$ and found that $A_{\omega}$ ranged from $0.38^{+0.4}_{-0.15}$ to $1.7^{+0.42}_{-0.37}$. These three studies measured clustering amplitudes in field studies.We find that the clustering amplitude in our Hot DOGs is somewhat higher than that observed in similar redshift studies conducted in the field.Specifically, at $\beta=0.8$, we find higher clustering  signal than in the SDF field studies \citep{2001Ouchi}, and weaker clustering at $\beta=0.6$, but it is similar clustering signal to the two Great Observatories Deep Origins Survey (GOODS) field studies \citep{2006Lee}. The rapid decrease in the clustering signal with decreasing $\beta$ values suggests that galaxies are more clustered at smaller angular separations.

\subsection{Overdensities around quasars and radio galaxies}\label{discussion}
  \begin{figure*}
\includegraphics[scale=0.58]{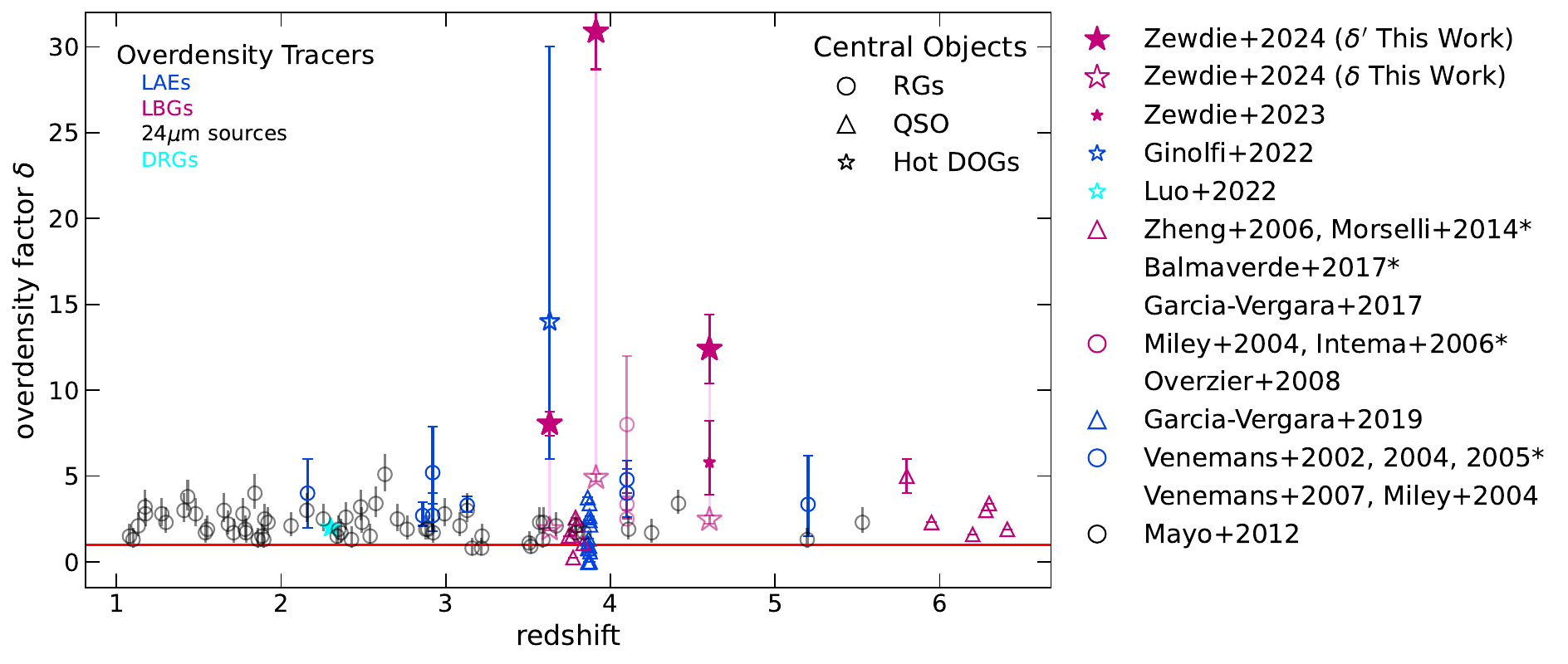}
\caption{Overdensity around high-redshift radio galaxies, quasars, and Hot DOGs as a function of redshift adapted from \protect\citetalias{2023Zewdie}. We added the three Hot DOGs overdensities ($\delta^{'}$ filled star and $\delta$ open star) that were found using the optimised selection criteria. The red horizontal line indicates a null overdensity. (*: In literature, overdensity was defined as $\delta = \frac{N_{found}}{N_{expected}} -1$, which we adjusted to align with the definition used in this study). }
\label{Overdensity_quasars}
\end{figure*}

\citetalias{2023Zewdie} conducted a comparison of the overdensities observed around Hot DOGs, quasars, and radio galaxies at different redshifts collected from the literature (see their Figure 12 and the discussion and references in their Section 4). The comparison encompasses various tracers such as LBGs around quasars \citep{2006Overzier, 2010Utsumi, 2014Morselli, 2017Balmaverde, 2017GarcaVergara, 2018Ota, 2020Mignoli}, and radio galaxies \citep{2004Miley, 2006Intema, 2008Overzier}; LAEs around quasars \citep{2019Garcia}, radio galaxies \citep{2002Venemans, 2004Venemans}, and Hot DOGs \citep{2022Ginolfi}. Additionally, the comparison includes red distant galaxies around a Hot DOG \citep{2022Luo} and 24 $\rm \mu m$ sources around radio galaxies \citep{2012Mayo}.

\cite{2004VOuchi} studied the properties of photometrically selected LBGs using deep SDF and SXDF imaging in \textit{R}-, \textit{i}-, \textit{z}-bands, finding that the selected LBGs had reasonably high completeness and low contamination from interlopers. \citetalias{2023Zewdie} modified the \cite{2004VOuchi} selection criteria by taking into account the filter curve and the IGM absorption model of \cite{1995Madau}, aiming to account for differences in the filters to ensure that we target the same sources as Ouchi et al. did, which is necessary for estimating the overdensity. \citetalias{2023Zewdie} also checked the modified selection criteria by overplotting different types of stars (including main sequence, giant, and supergiant stellar atmosphere models from \citealt{2004Castelli}, and M and L dwarfs from \citealt{2014Burgasser}) and low-redshift galaxy templates from \cite{1980Coleman}, shifted to redshifts from $z = 0$ to 3 (see Figures \ref{W0410_FLAG_color_simulated}-\ref{W2246_FLAG_color}). \citetalias{2023Zewdie} found no contamination from these interlopers, as shown in their Figure 4.

Here, we have updated their comparison by adding the overdensity factors around the three Hot DOGs studied in this work (see Figure \ref{Overdensity_quasars}), providing further evidence that Hot DOGs may live in some of the densest structures at their redshifts. For completeness, we show both the overdensity for the Hot DOG environments with ($\delta$) and without ($\delta^{\prime}$) correcting for contaminants (see section \ref{overdensity} for details). These dense environments may be related to their unique properties, such as extreme infrared luminosities and high SMBH accretion rates. Given the small sample size, we are not able to draw strong conclusions on the overdensity factor variation as a function of redshift.

\section{Conclusions}\label{conclusion}
In this paper, we studied the environments of three Hot DOGs by looking for companion LBGs using IMACS and GMOS-S photometry. In order to improve our sensitivity to LBGs, we have developed a novel process to optimise the photometric selection criteria using the COSMOS2020 combined catalog. Specifically, we used the HSC photometry of this field combined with its accurate photometric redshifts and SED classifications to adjust the colour selection criteria to target galaxies at the specific redshift of each Hot DOG we study. We summarize our results below.

\begin{enumerate} 
\item [1.] For the Hot DOG W0410-0916 at $z=3.631$ we find an overdensity of $\delta=1.83\pm 0.08$ when considering the whole FoV of the IMACS imaging $\rm (14.6 \times 14.6 ~Mpc^2)$ compared to the density of targets selected using the same criteria in the COSMOS2020 catalog. When focusing on the region within 2 Mpc of the Hot DOG, we find instead $\delta = 2.36 \pm 0.19$. When accounting for potential contaminants based on the redshifts and classifications of COSMOS2020, these overdensities increase to $\delta^{\prime} = 7.49 \pm 0.68$ and $12.38\pm 1.65$, respectively. Our results are consistent with the overdensity of LAEs of 14$^{+16}_{-8}$ found by \cite{2022Ginolfi} within a 0.4 Mpc radius of the Hot DOG.
\item[2.] For W0831+0140 at $z=3.912$, we also find an overdense field compared to COSMOS2020 with $\delta = 4.67 \pm 0.21$ within the entire IMACS FoV and $\delta = 4.38 \pm 0.33$ when focusing on region within 2 Mpc of the Hot DOG. When attempting to remove contaminants,  these estimates increase up to $\delta^{\prime} = 30.9\pm 2.0$ and $29.4\pm 3.84$, respectively.
\item[3.] We re-analized the GMOS-S observations presented by \citetalias{2023Zewdie} for W2246--0526 at $z=4.601$ to identify LBG companions using our method to optimise the selection criteria. We find an overdensity within the area of $\rm 4.7 \times 4.7 ~arcmin^2$ of $\delta = 2.5\pm 0.5$ that increases to $\delta^{\prime}=11.60\pm 1.96$ when attempting to remove contaminants. \citetalias{2023Zewdie}, instead, found an overdensity of $\delta = 5.8^{+2.4}_{-1.9}$ when using the selection criteria from \cite{2004VOuchi} and \cite{2006Yoshida} and comparing to the target density found in the SDF and SXDF. We find that while some of the difference could be explained by the different levels of contamination of the selection criteria and the different filters used, much of the difference may come from SDF/SXDF being underdense by a factor of $\sim$2 when compared to COSMOS2020. 
\item[4.] Analyzing the radial distribution of LBG candidates with respect to the Hot DOGs, we find that all three fields the overdensities are concentrated around the Hot DOGs (Figures \ref{distance_W0410}, \ref{distance_W0831}, and \ref{distance_W2246})
\end{enumerate}

We also compared our work with previous overdensity studies involving tracers such as LBGs, LAEs, and other companions around Hot DOGs, quasars, and radio galaxies. We found Hot DOGs may have some of the densest environments among luminous, active galaxies. Our results hence suggest that Hot DOGs are good tracers of dense proto-clusters. Additional spectroscopic follow-up observations are necessary to further constrain the properties of the environments of these objects and confirm whether Hot DOGs represent an early stage of formation for the BCGs found in the local Universe.

 \begin{acknowledgements}

We thank the anonymous referee for their constructive comments and suggestions, which improved this article. DZ acknowledges the support of ANID fellowship grants, grant No. 21211531. RJA was supported by FONDECYT grants number 191124 and 1231718, and by the ANID BASAL project FB210003. SIL is supported in part by the National Research Foundation (NRF) of South Africa (NRF Grant Number: 146053). Any opinion, finding, and conclusion or recommendation expressed in this material is that of the author(s), and the NRF does not accept any liability in this regard. JGL acknowledges support from "Programa de Inserción Académica 2024 Vicerrectoría Académica y Prorrectoría Pontificia Universidad Católica de Chile”. The work of PRME and DS was carried out at the Jet Propulsion Laboratory, California Institute of Technology, under a contract with NASA. TDS acknowledges the research project was supported by the Hellenic Foundation for Research and Innovation (HFRI) under the "2nd Call for HFRI Research Projects to support Faculty Members \& Researchers" (Project Number: 03382)

 \end{acknowledgements}

\bibliographystyle{aa} 
\bibliography{reference} 

\begin{thebibliography}{99}
\expandafter\ifx\csname natexlab\endcsname\relax\def\natexlab#1{#1}\fi

\bibitem[{{Angulo} {et~al.}(2012){Angulo}, {Springel}, {White}, {Cole}, {Jenkins}, {Baugh}, \& {Frenk}}]{2012Angulo}
{Angulo}, R.~E., {Springel}, V., {White}, S.~D.~M., {et~al.} 2012, \mnras, 425, 2722

\bibitem[{{Arnouts} {et~al.}(2002){Arnouts}, {Moscardini}, {Vanzella}, {Colombi}, {Cristiani}, {Fontana}, {Giallongo}, {Matarrese}, \& {Saracco}}]{2002Arnouts}
{Arnouts}, S., {Moscardini}, L., {Vanzella}, E., {et~al.} 2002, \mnras, 329, 355

\bibitem[{{Assef} {et~al.}(2022){Assef}, {Bauer}, {Blain}, {Brightman}, {D{\'\i}az-Santos}, {Eisenhardt}, {Jun}, {Stern}, {Tsai}, {Walton}, \& {Wu}}]{2022Assef}
{Assef}, R.~J., {Bauer}, F.~E., {Blain}, A.~W., {et~al.} 2022, \apj, 934, 101

\bibitem[{{Assef} {et~al.}(2020){Assef}, {Brightman}, {Walton}, {Stern}, {Bauer}, {Blain}, {D{\'\i}az-Santos}, {Eisenhardt}, {Hickox}, {Jun}, {Psychogyios}, {Tsai}, \& {Wu}}]{2020Assef}
{Assef}, R.~J., {Brightman}, M., {Walton}, D.~J., {et~al.} 2020, \apj, 897, 112

\bibitem[{{Assef} {et~al.}(2015){Assef}, {Eisenhardt}, {Stern}, {Tsai}, {Wu}, {Wylezalek}, {Blain}, {Bridge}, {Donoso}, {Gonzales}, {Griffith}, \& {Jarrett}}]{2015Assef}
{Assef}, R.~J., {Eisenhardt}, P.~R.~M., {Stern}, D., {et~al.} 2015, \apj, 804, 27

\bibitem[{{Assef} {et~al.}(2016){Assef}, {Walton}, {Brightman}, {Stern}, {Alexander}, {Bauer}, {Blain}, {Diaz-Santos}, {Eisenhardt}, {Finkelstein}, {Hickox}, {Tsai}, \& {Wu}}]{2016Assef}
{Assef}, R.~J., {Walton}, D.~J., {Brightman}, M., {et~al.} 2016, \apj, 819, 111

\bibitem[{{Ba{\~n}ados} {et~al.}(2013){Ba{\~n}ados}, {Venemans}, {Walter}, {Kurk}, {Overzier}, \& {Ouchi}}]{2013Banados}
{Ba{\~n}ados}, E., {Venemans}, B., {Walter}, F., {et~al.} 2013, \apj, 773, 178

\bibitem[{{Ba{\~n}ados} {et~al.}(2018){Ba{\~n}ados}, {Venemans}, {Mazzucchelli}, {Farina}, {Walter}, {Wang}, {Decarli}, {Stern}, {Fan}, {Davies}, {Hennawi}, {Simcoe}, {Turner}, {Rix}, {Yang}, {Kelson}, {Rudie}, \& {Winters}}]{2018Banados}
{Ba{\~n}ados}, E., {Venemans}, B.~P., {Mazzucchelli}, C., {et~al.} 2018, \nat, 553, 473

\bibitem[{{Balmaverde} {et~al.}(2017){Balmaverde}, {Gilli}, {Mignoli}, {Bolzonella}, {Brusa}, {Cappelluti}, {Comastri}, {Sani}, {Vanzella}, {Vignali}, {Vito}, \& {Zamorani}}]{2017Balmaverde}
{Balmaverde}, B., {Gilli}, R., {Mignoli}, M., {et~al.} 2017, \aap, 606, A23

\bibitem[{{Bertin} \& {Arnouts}(1996)}]{1996Bertin}
{Bertin}, E. \& {Arnouts}, S. 1996, \aaps, 117, 393

\bibitem[{{Bertin} {et~al.}(2002){Bertin}, {Mellier}, {Radovich}, {Missonnier}, {Didelon}, \& {Morin}}]{2002Bertin}
{Bertin}, E., {Mellier}, Y., {Radovich}, M., {et~al.} 2002, in Astronomical Society of the Pacific Conference Series, Vol. 281, Astronomical Data Analysis Software and Systems XI, ed. D.~A. {Bohlender}, D.~{Durand}, \& T.~H. {Handley}, 228

\bibitem[{{Bosman} {et~al.}(2020){Bosman}, {Kakiichi}, {Meyer}, {Gronke}, {Laporte}, \& {Ellis}}]{2020Bosman}
{Bosman}, S. E.~I., {Kakiichi}, K., {Meyer}, R.~A., {et~al.} 2020, \apj, 896, 49

\bibitem[{{Brammer} {et~al.}(2008){Brammer}, {van Dokkum}, \& {Coppi}}]{2008Brammer}
{Brammer}, G.~B., {van Dokkum}, P.~G., \& {Coppi}, P. 2008, \apj, 686, 1503

\bibitem[{{Burgasser}(2014)}]{2014Burgasser}
{Burgasser}, A.~J. 2014, in Astronomical Society of India Conference Series, Vol.~11, Astronomical Society of India Conference Series, 7--16

\bibitem[{{Castelli} \& {Kurucz}(2004)}]{2004Castelli}
{Castelli}, F. \& {Kurucz}, R.~L. 2004, \aap, 419, 725

\bibitem[{{Champagne} {et~al.}(2023){Champagne}, {Casey}, {Finkelstein}, {Bagley}, {Cooper}, {Larson}, {Long}, \& {Wang}}]{2023Champagne}
{Champagne}, J.~B., {Casey}, C.~M., {Finkelstein}, S.~L., {et~al.} 2023, \apj, 952, 99

\bibitem[{{Coleman} {et~al.}(1980){Coleman}, {Wu}, \& {Weedman}}]{1980Coleman}
{Coleman}, G.~D., {Wu}, C.~C., \& {Weedman}, D.~W. 1980, \apjs, 43, 393

\bibitem[{{Croom} {et~al.}(2005){Croom}, {Boyle}, {Shanks}, {Smith}, {Miller}, {Outram}, {Loaring}, {Hoyle}, \& {da {\^A}ngela}}]{2005Croom}
{Croom}, S.~M., {Boyle}, B.~J., {Shanks}, T., {et~al.} 2005, \mnras, 356, 415

\bibitem[{{Dayal} {et~al.}(2019){Dayal}, {Rossi}, {Shiralilou}, {Piana}, {Choudhury}, \& {Volonteri}}]{2019Dayal}
{Dayal}, P., {Rossi}, E.~M., {Shiralilou}, B., {et~al.} 2019, \mnras, 486, 2336

\bibitem[{{D{\'\i}az-Santos} {et~al.}(2018){D{\'\i}az-Santos}, {Assef}, {Blain}, {Aravena}, {Stern}, {Tsai}, {Eisenhardt}, {Wu}, {Jun}, {Dibert}, {Inami}, {Lansbury}, \& {Leclercq}}]{2018Tanio}
{D{\'\i}az-Santos}, T., {Assef}, R.~J., {Blain}, A.~W., {et~al.} 2018, Science, 362, 1034

\bibitem[{{D{\'\i}az-Santos} {et~al.}(2016){D{\'\i}az-Santos}, {Assef}, {Blain}, {Tsai}, {Aravena}, {Eisenhardt}, {Wu}, {Stern}, \& {Bridge}}]{2016Tanio}
{D{\'\i}az-Santos}, T., {Assef}, R.~J., {Blain}, A.~W., {et~al.} 2016, \apjl, 816, L6

\bibitem[{{Eisenhardt} {et~al.}(2012){Eisenhardt}, {Wu}, {Tsai}, {Assef}, {Benford}, {Blain}, {Bridge}, {Condon}, {Cushing}, {Cutri}, {Evans}, {Gelino}, {Griffith}, {Grillmair}, {Jarrett}, {Lonsdale}, {Masci}, {Mason}, {Petty}, {Sayers}, {Stanford}, {Stern}, {Wright}, \& {Yan}}]{2012Eisenhardt}
{Eisenhardt}, P. R.~M., {Wu}, J., {Tsai}, C.-W., {et~al.} 2012, \apj, 755, 173

\bibitem[{{Erben} {et~al.}(2005){Erben}, {Schirmer}, {Dietrich}, {Cordes}, {Haberzettl}, {Hetterscheidt}, {Hildebrandt}, {Schmithuesen}, {Schneider}, {Simon}, {Deul}, {Hook}, {Kaiser}, {Radovich}, {Benoist}, {Nonino}, {Olsen}, {Prandoni}, {Wichmann}, {Zaggia}, {Bomans}, {Dettmar}, \& {Miralles}}]{2005Erben}
{Erben}, T., {Schirmer}, M., {Dietrich}, J.~P., {et~al.} 2005, Astronomische Nachrichten, 326, 432

\bibitem[{{Fan} {et~al.}(2017){Fan}, {Jones}, {Han}, \& {Knudsen}}]{2017Fan}
{Fan}, L., {Jones}, S.~F., {Han}, Y., \& {Knudsen}, K.~K. 2017, \pasp, 129, 124101

\bibitem[{{Fan} {et~al.}(2023){Fan}, {Ba{\~n}ados}, \& {Simcoe}}]{2023Fan}
{Fan}, X., {Ba{\~n}ados}, E., \& {Simcoe}, R.~A. 2023, \araa, 61, 373

\bibitem[{{Finnerty} {et~al.}(2020){Finnerty}, {Larson}, {Soifer}, {Armus}, {Matthews}, {Jun}, {Moon}, {Melbourne}, {Gomez}, {Tsai}, {D{\'\i}az-Santos}, {Eisenhardt}, \& {Cushing}}]{2020Finnerty}
{Finnerty}, L., {Larson}, K., {Soifer}, B.~T., {et~al.} 2020, \apj, 905, 16

\bibitem[{{Gaia Collaboration} {et~al.}(2023){Gaia Collaboration}, {Vallenari}, {Brown}, {Prusti}, {de Bruijne}, {Arenou}, {Babusiaux}, {Biermann}, {Creevey}, {Ducourant}, {Evans}, {Eyer}, {Guerra}, {Hutton}, {Jordi}, {Klioner}, {Lammers}, {Lindegren}, {Luri}, {Mignard}, {Panem}, {Pourbaix}, {Randich}, {Sartoretti}, {Soubiran}, {Tanga}, {Walton}, {Bailer-Jones}, {Bastian}, {Drimmel}, {Jansen}, {Katz}, {Lattanzi}, {van Leeuwen}, {Bakker}, {Cacciari}, {Casta{\~n}eda}, {De Angeli}, {Fabricius}, {Fouesneau}, {Fr{\'e}mat}, {Galluccio}, {Guerrier}, {Heiter}, {Masana}, {Messineo}, {Mowlavi}, {Nicolas}, {Nienartowicz}, {Pailler}, {Panuzzo}, {Riclet}, {Roux}, {Seabroke}, {Sordo}, {Th{\'e}venin}, {Gracia-Abril}, {Portell}, {Teyssier}, {Altmann}, {Andrae}, {Audard}, {Bellas-Velidis}, {Benson}, {Berthier}, {Blomme}, {Burgess}, {Busonero}, {Busso}, {C{\'a}novas}, {Carry}, {Cellino}, {Cheek}, {Clementini}, {Damerdji}, {Davidson}, {de Teodoro}, {Nu{\~n}ez Campos}, {Delchambre}, {Dell'Oro}, {Esquej},
  {Fern{\'a}ndez-Hern{\'a}ndez}, {Fraile}, {Garabato}, {Garc{\'\i}a-Lario}, {Gosset}, {Haigron}, {Halbwachs}, {Hambly}, {Harrison}, {Hern{\'a}ndez}, {Hestroffer}, {Hodgkin}, {Holl}, {Jan{\ss}en}, {Jevardat de Fombelle}, {Jordan}, {Krone-Martins}, {Lanzafame}, {L{\"o}ffler}, {Marchal}, {Marrese}, {Moitinho}, {Muinonen}, {Osborne}, {Pancino}, {Pauwels}, {Recio-Blanco}, {Reyl{\'e}}, {Riello}, {Rimoldini}, {Roegiers}, {Rybizki}, {Sarro}, {Siopis}, {Smith}, {Sozzetti}, {Utrilla}, {van Leeuwen}, {Abbas}, {{\'A}brah{\'a}m}, {Abreu Aramburu}, {Aerts}, {Aguado}, {Ajaj}, {Aldea-Montero}, {Altavilla}, {{\'A}lvarez}, {Alves}, {Anders}, {Anderson}, {Anglada Varela}, {Antoja}, {Baines}, {Baker}, {Balaguer-N{\'u}{\~n}ez}, {Balbinot}, {Balog}, {Barache}, {Barbato}, {Barros}, {Barstow}, {Bartolom{\'e}}, {Bassilana}, {Bauchet}, {Becciani}, {Bellazzini}, {Berihuete}, {Bernet}, {Bertone}, {Bianchi}, {Binnenfeld}, {Blanco-Cuaresma}, {Blazere}, {Boch}, {Bombrun}, {Bossini}, {Bouquillon}, {Bragaglia}, {Bramante}, {Breedt},
  {Bressan}, {Brouillet}, {Brugaletta}, {Bucciarelli}, {Burlacu}, {Butkevich}, {Buzzi}, {Caffau}, {Cancelliere}, {Cantat-Gaudin}, {Carballo}, {Carlucci}, {Carnerero}, {Carrasco}, {Casamiquela}, {Castellani}, {Castro-Ginard}, {Chaoul}, {Charlot}, {Chemin}, {Chiaramida}, {Chiavassa}, {Chornay}, {Comoretto}, {Contursi}, {Cooper}, {Cornez}, {Cowell}, {Crifo}, {Cropper}, {Crosta}, {Crowley}, {Dafonte}, {Dapergolas}, {David}, {David}, {de Laverny}, {De Luise}, {De March}, {De Ridder}, {de Souza}, {de Torres}, {del Peloso}, {del Pozo}, {Delbo}, {Delgado}, {Delisle}, {Demouchy}, {Dharmawardena}, {Di Matteo}, {Diakite}, {Diener}, {Distefano}, {Dolding}, {Edvardsson}, {Enke}, {Fabre}, {Fabrizio}, {Faigler}, {Fedorets}, {Fernique}, {Fienga}, {Figueras}, {Fournier}, {Fouron}, {Fragkoudi}, {Gai}, {Garcia-Gutierrez}, {Garcia-Reinaldos}, {Garc{\'\i}a-Torres}, {Garofalo}, {Gavel}, {Gavras}, {Gerlach}, {Geyer}, {Giacobbe}, {Gilmore}, {Girona}, {Giuffrida}, {Gomel}, {Gomez}, {Gonz{\'a}lez-N{\'u}{\~n}ez},
  {Gonz{\'a}lez-Santamar{\'\i}a}, {Gonz{\'a}lez-Vidal}, {Granvik}, {Guillout}, {Guiraud}, {Guti{\'e}rrez-S{\'a}nchez}, {Guy}, {Hatzidimitriou}, {Hauser}, {Haywood}, {Helmer}, {Helmi}, {Sarmiento}, {Hidalgo}, {Hilger}, {H{\l}adczuk}, {Hobbs}, {Holland}, {Huckle}, {Jardine}, {Jasniewicz}, {Jean-Antoine Piccolo}, {Jim{\'e}nez-Arranz}, {Jorissen}, {Juaristi Campillo}, {Julbe}, {Karbevska}, {Kervella}, {Khanna}, {Kontizas}, {Kordopatis}, {Korn}, {K{\'o}sp{\'a}l}, {Kostrzewa-Rutkowska}, {Kruszy{\'n}ska}, {Kun}, {Laizeau}, {Lambert}, {Lanza}, {Lasne}, {Le Campion}, {Lebreton}, {Lebzelter}, {Leccia}, {Leclerc}, {Lecoeur-Taibi}, {Liao}, {Licata}, {Lindstr{\o}m}, {Lister}, {Livanou}, {Lobel}, {Lorca}, {Loup}, {Madrero Pardo}, {Magdaleno Romeo}, {Managau}, {Mann}, {Manteiga}, {Marchant}, {Marconi}, {Marcos}, {Marcos Santos}, {Mar{\'\i}n Pina}, {Marinoni}, {Marocco}, {Marshall}, {Martin Polo}, {Mart{\'\i}n-Fleitas}, {Marton}, {Mary}, {Masip}, {Massari}, {Mastrobuono-Battisti}, {Mazeh}, {McMillan}, {Messina}, {Michalik},
  {Millar}, {Mints}, {Molina}, {Molinaro}, {Moln{\'a}r}, {Monari}, {Mongui{\'o}}, {Montegriffo}, {Montero}, {Mor}, {Mora}, {Morbidelli}, {Morel}, {Morris}, {Muraveva}, {Murphy}, {Musella}, {Nagy}, {Noval}, {Oca{\~n}a}, {Ogden}, {Ordenovic}, {Osinde}, {Pagani}, {Pagano}, {Palaversa}, {Palicio}, {Pallas-Quintela}, {Panahi}, {Payne-Wardenaar}, {Pe{\~n}alosa Esteller}, {Penttil{\"a}}, {Pichon}, {Piersimoni}, {Pineau}, {Plachy}, {Plum}, {Poggio}, {Pr{\v{s}}a}, {Pulone}, {Racero}, {Ragaini}, {Rainer}, {Raiteri}, {Rambaux}, {Ramos}, {Ramos-Lerate}, {Re Fiorentin}, {Regibo}, {Richards}, {Rios Diaz}, {Ripepi}, {Riva}, {Rix}, {Rixon}, {Robichon}, {Robin}, {Robin}, {Roelens}, {Rogues}, {Rohrbasser}, {Romero-G{\'o}mez}, {Rowell}, {Royer}, {Ruz Mieres}, {Rybicki}, {Sadowski}, {S{\'a}ez N{\'u}{\~n}ez}, {Sagrist{\`a} Sell{\'e}s}, {Sahlmann}, {Salguero}, {Samaras}, {Sanchez Gimenez}, {Sanna}, {Santove{\~n}a}, {Sarasso}, {Schultheis}, {Sciacca}, {Segol}, {Segovia}, {S{\'e}gransan}, {Semeux}, {Shahaf}, {Siddiqui}, {Siebert},
  {Siltala}, {Silvelo}, {Slezak}, {Slezak}, {Smart}, {Snaith}, {Solano}, {Solitro}, {Souami}, {Souchay}, {Spagna}, {Spina}, {Spoto}, {Steele}, {Steidelm{\"u}ller}, {Stephenson}, {S{\"u}veges}, {Surdej}, {Szabados}, {Szegedi-Elek}, {Taris}, {Taylor}, {Teixeira}, {Tolomei}, {Tonello}, {Torra}, {Torra}, {Torralba Elipe}, {Trabucchi}, {Tsounis}, {Turon}, {Ulla}, {Unger}, {Vaillant}, {van Dillen}, {van Reeven}, {Vanel}, {Vecchiato}, {Viala}, {Vicente}, {Voutsinas}, {Weiler}, {Wevers}, {Wyrzykowski}, {Yoldas}, {Yvard}, {Zhao}, {Zorec}, {Zucker}, \& {Zwitter}}]{2023Gaia}
{Gaia Collaboration}, {Vallenari}, A., {Brown}, A.~G.~A., {et~al.} 2023, \aap, 674, A1

\bibitem[{{Garc{\'\i}a-Vergara} {et~al.}(2019){Garc{\'\i}a-Vergara}, {Hennawi}, {Barrientos}, \& {Arrigoni Battaia}}]{2019Garcia}
{Garc{\'\i}a-Vergara}, C., {Hennawi}, J.~F., {Barrientos}, L.~F., \& {Arrigoni Battaia}, F. 2019, \apj, 886, 79

\bibitem[{{Garc{\'\i}a-Vergara} {et~al.}(2017){Garc{\'\i}a-Vergara}, {Hennawi}, {Barrientos}, \& {Rix}}]{2017GarcaVergara}
{Garc{\'\i}a-Vergara}, C., {Hennawi}, J.~F., {Barrientos}, L.~F., \& {Rix}, H.-W. 2017, \apj, 848, 7

\bibitem[{{Ginolfi} {et~al.}(2022){Ginolfi}, {Piconcelli}, {Zappacosta}, {Jones}, {Pentericci}, {Maiolino}, {Travascio}, {Menci}, {Carniani}, {Rizzo}, {Arrigoni Battaia}, {Cantalupo}, {De Breuck}, {Graziani}, {Knudsen}, {Laursen}, {Mainieri}, {Schneider}, {Stanley}, {Valiante}, \& {Verhamme}}]{2022Ginolfi}
{Ginolfi}, M., {Piconcelli}, E., {Zappacosta}, L., {et~al.} 2022, Nature Communications, 13, 4574

\bibitem[{{Harikane} {et~al.}(2016){Harikane}, {Ouchi}, {Ono}, {More}, {Saito}, {Lin}, {Coupon}, {Shimasaku}, {Shibuya}, {Price}, {Lin}, {Hsieh}, {Ishigaki}, {Komiyama}, {Silverman}, {Takata}, {Tamazawa}, \& {Toshikawa}}]{2016Harikane}
{Harikane}, Y., {Ouchi}, M., {Ono}, Y., {et~al.} 2016, \apj, 821, 123

\bibitem[{{Hoaglin} {et~al.}(1983){Hoaglin}, {Mosteller}, \& {Tukey}}]{1983Hoaglin}
{Hoaglin}, D.~C., {Mosteller}, F., \& {Tukey}, J.~W. 1983, {Understanding robust and exploratory data anlysis}

\bibitem[{{Husband} {et~al.}(2013){Husband}, {Bremer}, {Stanway}, {Davies}, {Lehnert}, \& {Douglas}}]{2013Husband}
{Husband}, K., {Bremer}, M.~N., {Stanway}, E.~R., {et~al.} 2013, \mnras, 432, 2869

\bibitem[{{Ilbert} {et~al.}(2006){Ilbert}, {Arnouts}, {McCracken}, {Bolzonella}, {Bertin}, {Le F{\`e}vre}, {Mellier}, {Zamorani}, {Pell{\`o}}, {Iovino}, {Tresse}, {Le Brun}, {Bottini}, {Garilli}, {Maccagni}, {Picat}, {Scaramella}, {Scodeggio}, {Vettolani}, {Zanichelli}, {Adami}, {Bardelli}, {Cappi}, {Charlot}, {Ciliegi}, {Contini}, {Cucciati}, {Foucaud}, {Franzetti}, {Gavignaud}, {Guzzo}, {Marano}, {Marinoni}, {Mazure}, {Meneux}, {Merighi}, {Paltani}, {Pollo}, {Pozzetti}, {Radovich}, {Zucca}, {Bondi}, {Bongiorno}, {Busarello}, {de La Torre}, {Gregorini}, {Lamareille}, {Mathez}, {Merluzzi}, {Ripepi}, {Rizzo}, \& {Vergani}}]{2006Ilbert}
{Ilbert}, O., {Arnouts}, S., {McCracken}, H.~J., {et~al.} 2006, \aap, 457, 841

\bibitem[{{Inayoshi} {et~al.}(2020){Inayoshi}, {Visbal}, \& {Haiman}}]{2020Inayoshi}
{Inayoshi}, K., {Visbal}, E., \& {Haiman}, Z. 2020, \araa, 58, 27

\bibitem[{{Intema} {et~al.}(2006){Intema}, {Venemans}, {Kurk}, {Ouchi}, {Kodama}, {R{\"o}ttgering}, {Miley}, \& {Overzier}}]{2006Intema}
{Intema}, H.~T., {Venemans}, B.~P., {Kurk}, J.~D., {et~al.} 2006, \aap, 456, 433

\bibitem[{{Jones} {et~al.}(2017){Jones}, {Blain}, {Assef}, {Eisenhardt}, {Lonsdale}, {Condon}, {Farrah}, {Tsai}, {Bridge}, {Wu}, {Wright}, \& {Jarrett}}]{2017Jones}
{Jones}, S.~F., {Blain}, A.~W., {Assef}, R.~J., {et~al.} 2017, \mnras, 469, 4565

\bibitem[{{Jones} {et~al.}(2014){Jones}, {Blain}, {Stern}, {Assef}, {Bridge}, {Eisenhardt}, {Petty}, {Wu}, {Tsai}, {Cutri}, {Wright}, \& {Yan}}]{2014Jones}
{Jones}, S.~F., {Blain}, A.~W., {Stern}, D., {et~al.} 2014, \mnras, 443, 146

\bibitem[{{Jun} {et~al.}(2020){Jun}, {Assef}, {Bauer}, {Blain}, {D{\'\i}az-Santos}, {Eisenhardt}, {Stern}, {Tsai}, {Wright}, \& {Wu}}]{2020Jun}
{Jun}, H.~D., {Assef}, R.~J., {Bauer}, F.~E., {et~al.} 2020, \apj, 888, 110

\bibitem[{{Kashikawa} {et~al.}(2007){Kashikawa}, {Kitayama}, {Doi}, {Misawa}, {Komiyama}, \& {Ota}}]{2007Kashikawa}
{Kashikawa}, N., {Kitayama}, T., {Doi}, M., {et~al.} 2007, \apj, 663, 765

\bibitem[{{Kim} {et~al.}(2009){Kim}, {Stiavelli}, {Trenti}, {Pavlovsky}, {Djorgovski}, {Scarlata}, {Stern}, {Mahabal}, {Thompson}, {Dickinson}, {Panagia}, \& {Meylan}}]{2009Kim}
{Kim}, S., {Stiavelli}, M., {Trenti}, M., {et~al.} 2009, \apj, 695, 809

\bibitem[{{Lambert} {et~al.}(2024){Lambert}, {Assef}, {Mazzucchelli}, {Ba{\~n}ados}, {Aravena}, {Barrientos}, {Gonz{\'a}lez-L{\'o}pez}, {Hu}, {Infante}, {Malhotra}, {Moya-Sierralta}, {Rhoads}, {Valdes}, {Wang}, {Wold}, \& {Zheng}}]{2024Lambert}
{Lambert}, T.~S., {Assef}, R.~J., {Mazzucchelli}, C., {et~al.} 2024, arXiv e-prints, arXiv:2402.06870

\bibitem[{{Landy} \& {Szalay}(1993)}]{1993Landy}
{Landy}, S.~D. \& {Szalay}, A.~S. 1993, \apj, 412, 64

\bibitem[{{Lee} {et~al.}(2006){Lee}, {Giavalisco}, {Gnedin}, {Somerville}, {Ferguson}, {Dickinson}, \& {Ouchi}}]{2006Lee}
{Lee}, K.-S., {Giavalisco}, M., {Gnedin}, O.~Y., {et~al.} 2006, \apj, 642, 63

\bibitem[{{Li} {et~al.}(2024){Li}, {Assef}, {Tsai}, {Wu}, {Eisenhardt}, {Stern}, {D{\'\i}az-Santos}, {Blain}, {Jun}, {Fern{\'a}ndez Arand{\'a}}, \& {Zewdie}}]{2024Li}
{Li}, G., {Assef}, R.~J., {Tsai}, C.-W., {et~al.} 2024, arXiv e-prints, arXiv:2405.20479

\bibitem[{{Li} {et~al.}(2007){Li}, {Hernquist}, {Robertson}, {Cox}, {Hopkins}, {Springel}, {Gao}, {Di Matteo}, {Zentner}, {Jenkins}, \& {Yoshida}}]{2007Li}
{Li}, Y., {Hernquist}, L., {Robertson}, B., {et~al.} 2007, \apj, 665, 187

\bibitem[{{Luo} {et~al.}(2022){Luo}, {Fan}, {Zou}, {Shen}, {Lin}, {Hu}, {Lin}, {Tao}, \& {Chen}}]{2022Luo}
{Luo}, Y., {Fan}, L., {Zou}, H., {et~al.} 2022, \apj, 935, 80

\bibitem[{{Madau}(1995)}]{1995Madau}
{Madau}, P. 1995, \apj, 441, 18

\bibitem[{{Mayo} {et~al.}(2012){Mayo}, {Vernet}, {De Breuck}, {Galametz}, {Seymour}, \& {Stern}}]{2012Mayo}
{Mayo}, J.~H., {Vernet}, J., {De Breuck}, C., {et~al.} 2012, \aap, 539, A33

\bibitem[{{Mazzucchelli} {et~al.}(2017){Mazzucchelli}, {Ba{\~n}ados}, {Decarli}, {Farina}, {Venemans}, {Walter}, \& {Overzier}}]{2017Mazzucchelli}
{Mazzucchelli}, C., {Ba{\~n}ados}, E., {Decarli}, R., {et~al.} 2017, \apj, 834, 83

\bibitem[{{Mignoli} {et~al.}(2020){Mignoli}, {Gilli}, {Decarli}, {Vanzella}, {Balmaverde}, {Cappelluti}, {Cassar{\`a}}, {Comastri}, {Cusano}, {Iwasawa}, {Marchesi}, {Prandoni}, {Vignali}, {Vito}, {Zamorani}, {Chiaberge}, \& {Norman}}]{2020Mignoli}
{Mignoli}, M., {Gilli}, R., {Decarli}, R., {et~al.} 2020, \aap, 642, L1

\bibitem[{{Miley} {et~al.}(2004){Miley}, {Overzier}, {Tsvetanov}, {Bouwens}, {Ben{\'\i}tez}, {Blakeslee}, {Ford}, {Illingworth}, {Postman}, {Rosati}, {Clampin}, {Hartig}, {Zirm}, {R{\"o}ttgering}, {Venemans}, {Ardila}, {Bartko}, {Broadhurst}, {Brown}, {Burrows}, {Cheng}, {Cross}, {De Breuck}, {Feldman}, {Franx}, {Golimowski}, {Gronwall}, {Infante}, {Martel}, {Menanteau}, {Meurer}, {Sirianni}, {Kimble}, {Krist}, {Sparks}, {Tran}, {White}, \& {Zheng}}]{2004Miley}
{Miley}, G.~K., {Overzier}, R.~A., {Tsvetanov}, Z.~I., {et~al.} 2004, \nat, 427, 47

\bibitem[{{Morselli} {et~al.}(2014){Morselli}, {Mignoli}, {Gilli}, {Vignali}, {Comastri}, {Sani}, {Cappelluti}, {Zamorani}, {Brusa}, {Gallozzi}, \& {Vanzella}}]{2014Morselli}
{Morselli}, L., {Mignoli}, M., {Gilli}, R., {et~al.} 2014, \aap, 568, A1

\bibitem[{{Noirot} {et~al.}(2018){Noirot}, {Stern}, {Mei}, {Wylezalek}, {Cooke}, {De Breuck}, {Galametz}, {Hatch}, {Vernet}, {Brodwin}, {Eisenhardt}, {Gonzalez}, {Jarvis}, {Rettura}, {Seymour}, \& {Stanford}}]{2018Noirot}
{Noirot}, G., {Stern}, D., {Mei}, S., {et~al.} 2018, \apj, 859, 38

\bibitem[{{Ota} {et~al.}(2018){Ota}, {Venemans}, {Taniguchi}, {Kashikawa}, {Nakata}, {Harikane}, {Ba{\~n}ados}, {Overzier}, {Riechers}, {Walter}, {Toshikawa}, {Shibuya}, \& {Jiang}}]{2018Ota}
{Ota}, K., {Venemans}, B.~P., {Taniguchi}, Y., {et~al.} 2018, \apj, 856, 109

\bibitem[{{Ouchi} {et~al.}(2001){Ouchi}, {Shimasaku}, {Okamura}, {Doi}, {Furusawa}, {Hamabe}, {Kimura}, {Komiyama}, {Miyazaki}, {Miyazaki}, {Nakata}, {Sekiguchi}, {Yagi}, \& {Yasuda}}]{2001Ouchi}
{Ouchi}, M., {Shimasaku}, K., {Okamura}, S., {et~al.} 2001, \apjl, 558, L83

\bibitem[{{Ouchi} {et~al.}(2004){Ouchi}, {Shimasaku}, {Okamura}, {Furusawa}, {Kashikawa}, {Ota}, {Doi}, {Hamabe}, {Kimura}, {Komiyama}, {Miyazaki}, {Miyazaki}, {Nakata}, {Sekiguchi}, {Yagi}, \& {Yasuda}}]{2004VOuchi}
{Ouchi}, M., {Shimasaku}, K., {Okamura}, S., {et~al.} 2004, \apj, 611, 660

\bibitem[{{Overzier} {et~al.}(2008){Overzier}, {Bouwens}, {Cross}, {Venemans}, {Miley}, {Zirm}, {Ben{\'\i}tez}, {Blakeslee}, {Coe}, {Demarco}, {Ford}, {Homeier}, {Illingworth}, {Kurk}, {Martel}, {Mei}, {Oliveira}, {R{\"o}ttgering}, {Tsvetanov}, \& {Zheng}}]{2008Overzier}
{Overzier}, R.~A., {Bouwens}, R.~J., {Cross}, N.~J.~G., {et~al.} 2008, \apj, 673, 143

\bibitem[{{Overzier} {et~al.}(2006{\natexlab{a}}){Overzier}, {Bouwens}, {Illingworth}, \& {Franx}}]{2006OverzierC}
{Overzier}, R.~A., {Bouwens}, R.~J., {Illingworth}, G.~D., \& {Franx}, M. 2006{\natexlab{a}}, \apjl, 648, L5

\bibitem[{{Overzier} {et~al.}(2009){Overzier}, {Guo}, {Kauffmann}, {De Lucia}, {Bouwens}, \& {Lemson}}]{2009Overzier}
{Overzier}, R.~A., {Guo}, Q., {Kauffmann}, G., {et~al.} 2009, \mnras, 394, 577

\bibitem[{{Overzier} {et~al.}(2006{\natexlab{b}}){Overzier}, {Miley}, {Bouwens}, {Cross}, {Zirm}, {Ben{\'\i}tez}, {Blakeslee}, {Clampin}, {Demarco}, {Ford}, {Hartig}, {Illingworth}, {Martel}, {R{\"o}ttgering}, {Venemans}, {Ardila}, {Bartko}, {Bradley}, {Broadhurst}, {Coe}, {Feldman}, {Franx}, {Golimowski}, {Goto}, {Gronwall}, {Holden}, {Homeier}, {Infante}, {Kimble}, {Krist}, {Mei}, {Menanteau}, {Meurer}, {Motta}, {Postman}, {Rosati}, {Sirianni}, {Sparks}, {Tran}, {Tsvetanov}, {White}, \& {Zheng}}]{2006Overzier}
{Overzier}, R.~A., {Miley}, G.~K., {Bouwens}, R.~J., {et~al.} 2006{\natexlab{b}}, \apj, 637, 58

\bibitem[{{Roche} \& {Eales}(1999)}]{1999Roche}
{Roche}, N. \& {Eales}, S.~A. 1999, \mnras, 307, 703

\bibitem[{{Schirmer}(2013)}]{2013Schirmer}
{Schirmer}, M. 2013, \apjs, 209, 21

\bibitem[{{Scoville} {et~al.}(2007){Scoville}, {Aussel}, {Brusa}, {Capak}, {Carollo}, {Elvis}, {Giavalisco}, {Guzzo}, {Hasinger}, {Impey}, {Kneib}, {LeFevre}, {Lilly}, {Mobasher}, {Renzini}, {Rich}, {Sanders}, {Schinnerer}, {Schminovich}, {Shopbell}, {Taniguchi}, \& {Tyson}}]{2007Scoville}
{Scoville}, N., {Aussel}, H., {Brusa}, M., {et~al.} 2007, \apjs, 172, 1

\bibitem[{{Shapley} {et~al.}(2003){Shapley}, {Steidel}, {Pettini}, \& {Adelberger}}]{2003Shapley}
{Shapley}, A.~E., {Steidel}, C.~C., {Pettini}, M., \& {Adelberger}, K.~L. 2003, \apj, 588, 65

\bibitem[{{Soltan}(1982)}]{1982Soltan}
{Soltan}, A. 1982, \mnras, 200, 115

\bibitem[{{Steidel} {et~al.}(2003){Steidel}, {Adelberger}, {Shapley}, {Pettini}, {Dickinson}, \& {Giavalisco}}]{2003Steidel}
{Steidel}, C.~C., {Adelberger}, K.~L., {Shapley}, A.~E., {et~al.} 2003, \apj, 592, 728

\bibitem[{{Stern} {et~al.}(2014){Stern}, {Lansbury}, {Assef}, {Brandt}, {Alexander}, {Ballantyne}, {Balokovi{\'c}}, {Bauer}, {Benford}, {Blain}, {Boggs}, {Bridge}, {Brightman}, {Christensen}, {Comastri}, {Craig}, {Del Moro}, {Eisenhardt}, {Gandhi}, {Griffith}, {Hailey}, {Harrison}, {Hickox}, {Jarrett}, {Koss}, {Lake}, {LaMassa}, {Luo}, {Tsai}, {Urry}, {Walton}, {Wright}, {Wu}, {Yan}, \& {Zhang}}]{2014Stern}
{Stern}, D., {Lansbury}, G.~B., {Assef}, R.~J., {et~al.} 2014, \apj, 794, 102

\bibitem[{{Tachibana} \& {Miller}(2018)}]{2018Tachibana}
{Tachibana}, Y. \& {Miller}, A.~A. 2018, \pasp, 130, 128001

\bibitem[{{Tonry} {et~al.}(2012){Tonry}, {Stubbs}, {Lykke}, {Doherty}, {Shivvers}, {Burgett}, {Chambers}, {Hodapp}, {Kaiser}, {Kudritzki}, {Magnier}, {Morgan}, {Price}, \& {Wainscoat}}]{2012Tonry}
{Tonry}, J.~L., {Stubbs}, C.~W., {Lykke}, K.~R., {et~al.} 2012, \apj, 750, 99

\bibitem[{{Tsai} {et~al.}(2018){Tsai}, {Eisenhardt}, {Jun}, {Wu}, {Assef}, {Blain}, {D{\'\i}az-Santos}, {Jones}, {Stern}, {Wright}, \& {Yeh}}]{2018Tsai}
{Tsai}, C.-W., {Eisenhardt}, P. R.~M., {Jun}, H.~D., {et~al.} 2018, \apj, 868, 15

\bibitem[{{Tsai} {et~al.}(2015){Tsai}, {Eisenhardt}, {Wu}, {Stern}, {Assef}, {Blain}, {Bridge}, {Benford}, {Cutri}, {Griffith}, {Jarrett}, {Lonsdale}, {Masci}, {Moustakas}, {Petty}, {Sayers}, {Stanford}, {Wright}, {Yan}, {Leisawitz}, {Liu}, {Mainzer}, {McLean}, {Padgett}, {Skrutskie}, {Gelino}, {Beichman}, \& {Juneau}}]{2015Tsai}
{Tsai}, C.-W., {Eisenhardt}, P. R.~M., {Wu}, J., {et~al.} 2015, \apj, 805, 90

\bibitem[{{Uchiyama} {et~al.}(2018){Uchiyama}, {Toshikawa}, {Kashikawa}, {Overzier}, {Chiang}, {Marinello}, {Tanaka}, {Niino}, {Ishikawa}, {Onoue}, {Ichikawa}, {Akiyama}, {Coupon}, {Harikane}, {Imanishi}, {Kodama}, {Komiyama}, {Lee}, {Lin}, {Miyazaki}, {Nagao}, {Nishizawa}, {Ono}, {Ouchi}, \& {Wang}}]{2018Uchiyama}
{Uchiyama}, H., {Toshikawa}, J., {Kashikawa}, N., {et~al.} 2018, \pasj, 70, S32

\bibitem[{{Utsumi} {et~al.}(2010){Utsumi}, {Goto}, {Kashikawa}, {Miyazaki}, {Komiyama}, {Furusawa}, \& {Overzier}}]{2010Utsumi}
{Utsumi}, Y., {Goto}, T., {Kashikawa}, N., {et~al.} 2010, \apj, 721, 1680

\bibitem[{{Venemans} {et~al.}(2002){Venemans}, {Kurk}, {Miley}, {R{\"o}ttgering}, {van Breugel}, {Carilli}, {De Breuck}, {Ford}, {Heckman}, {McCarthy}, \& {Pentericci}}]{2002Venemans}
{Venemans}, B.~P., {Kurk}, J.~D., {Miley}, G.~K., {et~al.} 2002, \apjl, 569, L11

\bibitem[{{Venemans} {et~al.}(2007){Venemans}, {R{\"o}ttgering}, {Miley}, {van Breugel}, {de Breuck}, {Kurk}, {Pentericci}, {Stanford}, {Overzier}, {Croft}, \& {Ford}}]{2007Venemans}
{Venemans}, B.~P., {R{\"o}ttgering}, H.~J.~A., {Miley}, G.~K., {et~al.} 2007, \aap, 461, 823

\bibitem[{{Venemans} {et~al.}(2004){Venemans}, {R{\"o}ttgering}, {Overzier}, {Miley}, {De Breuck}, {Kurk}, {van Breugel}, {Carilli}, {Ford}, {Heckman}, {McCarthy}, \& {Pentericci}}]{2004Venemans}
{Venemans}, B.~P., {R{\"o}ttgering}, H.~J.~A., {Overzier}, R.~A., {et~al.} 2004, \aap, 424, L17

\bibitem[{{Wang} {et~al.}(2021){Wang}, {Yang}, {Fan}, {Hennawi}, {Barth}, {Banados}, {Bian}, {Boutsia}, {Connor}, {Davies}, {Decarli}, {Eilers}, {Farina}, {Green}, {Jiang}, {Li}, {Mazzucchelli}, {Nanni}, {Schindler}, {Venemans}, {Walter}, {Wu}, \& {Yue}}]{2021Wang}
{Wang}, F., {Yang}, J., {Fan}, X., {et~al.} 2021, \apjl, 907, L1

\bibitem[{{Weaver} {et~al.}(2022){Weaver}, {Kauffmann}, {Ilbert}, {McCracken}, {Moneti}, {Toft}, {Brammer}, {Shuntov}, {Davidzon}, {Hsieh}, {Laigle}, {Anastasiou}, {Jespersen}, {Vinther}, {Capak}, {Casey}, {McPartland}, {Milvang-Jensen}, {Mobasher}, {Sanders}, {Zalesky}, {Arnouts}, {Aussel}, {Dunlop}, {Faisst}, {Franx}, {Furtak}, {Fynbo}, {Gould}, {Greve}, {Gwyn}, {Kartaltepe}, {Kashino}, {Koekemoer}, {Kokorev}, {Le F{\`e}vre}, {Lilly}, {Masters}, {Magdis}, {Mehta}, {Peng}, {Riechers}, {Salvato}, {Sawicki}, {Scarlata}, {Scoville}, {Shirley}, {Silverman}, {Sneppen}, {Smolc̆i{\'c}}, {Steinhardt}, {Stern}, {Tanaka}, {Taniguchi}, {Teplitz}, {Vaccari}, {Wang}, \& {Zamorani}}]{2022Weaver}
{Weaver}, J.~R., {Kauffmann}, O.~B., {Ilbert}, O., {et~al.} 2022, \apjs, 258, 11

\bibitem[{{Wright} {et~al.}(2010){Wright}, {Eisenhardt}, {Mainzer}, {Ressler}, {Cutri}, {Jarrett}, {Kirkpatrick}, {Padgett}, {McMillan}, {Skrutskie}, {Stanford}, {Cohen}, {Walker}, {Mather}, {Leisawitz}, {Gautier}, {McLean}, {Benford}, {Lonsdale}, {Blain}, {Mendez}, {Irace}, {Duval}, {Liu}, {Royer}, {Heinrichsen}, {Howard}, {Shannon}, {Kendall}, {Walsh}, {Larsen}, {Cardon}, {Schick}, {Schwalm}, {Abid}, {Fabinsky}, {Naes}, \& {Tsai}}]{2010Wright}
{Wright}, E.~L., {Eisenhardt}, P. R.~M., {Mainzer}, A.~K., {et~al.} 2010, \aj, 140, 1868

\bibitem[{{Wu} {et~al.}(2012){Wu}, {Tsai}, {Sayers}, {Benford}, {Bridge}, {Blain}, {Eisenhardt}, {Stern}, {Petty}, {Assef}, {Bussmann}, {Comerford}, {Cutri}, {Evans}, {Griffith}, {Jarrett}, {Lake}, {Lonsdale}, {Rho}, {Stanford}, {Weiner}, {Wright}, \& {Yan}}]{2012Wu}
{Wu}, J., {Tsai}, C.-W., {Sayers}, J., {et~al.} 2012, \apj, 756, 96

\bibitem[{{Wylezalek} {et~al.}(2013){Wylezalek}, {Galametz}, {Stern}, {Vernet}, {De Breuck}, {Seymour}, {Brodwin}, {Eisenhardt}, {Gonzalez}, {Hatch}, {Jarvis}, {Rettura}, {Stanford}, \& {Stevens}}]{2013Wylezalek}
{Wylezalek}, D., {Galametz}, A., {Stern}, D., {et~al.} 2013, \apj, 769, 79

\bibitem[{{Yoshida} {et~al.}(2006){Yoshida}, {Shimasaku}, {Kashikawa}, {Ouchi}, {Okamura}, {Ajiki}, {Akiyama}, {Ando}, {Aoki}, {Doi}, {Furusawa}, {Hayashino}, {Iwamuro}, {Iye}, {Karoji}, {Kobayashi}, {Kodaira}, {Kodama}, {Komiyama}, {Malkan}, {Matsuda}, {Miyazaki}, {Mizumoto}, {Morokuma}, {Motohara}, {Murayama}, {Nagao}, {Nariai}, {Ohta}, {Sasaki}, {Sato}, {Sekiguchi}, {Shioya}, {Tamura}, {Taniguchi}, {Umemura}, {Yamada}, \& {Yasuda}}]{2006Yoshida}
{Yoshida}, M., {Shimasaku}, K., {Kashikawa}, N., {et~al.} 2006, \apj, 653, 988

\bibitem[{{Zewdie} {et~al.}(2023){Zewdie}, {Assef}, {Mazzucchelli}, {Aravena}, {Blain}, {D{\'\i}az-Santos}, {Eisenhardt}, {Jun}, {Stern}, {Tsai}, \& {Wu}}]{2023Zewdie}
{Zewdie}, D., {Assef}, R.~J., {Mazzucchelli}, C., {et~al.} 2023, \aap, 677, A54

\bibitem[{{Zheng} {et~al.}(2006){Zheng}, {Overzier}, {Bouwens}, {White}, {Ford}, {Ben{\'\i}tez}, {Blakeslee}, {Bradley}, {Jee}, {Martel}, {Mei}, {Zirm}, {Illingworth}, {Clampin}, {Hartig}, {Ardila}, {Bartko}, {Broadhurst}, {Brown}, {Burrows}, {Cheng}, {Cross}, {Demarco}, {Feldman}, {Franx}, {Golimowski}, {Goto}, {Gronwall}, {Holden}, {Homeier}, {Infante}, {Kimble}, {Krist}, {Lesser}, {Menanteau}, {Meurer}, {Miley}, {Motta}, {Postman}, {Rosati}, {Sirianni}, {Sparks}, {Tran}, \& {Tsvetanov}}]{2006Zheng}
{Zheng}, W., {Overzier}, R.~A., {Bouwens}, R.~J., {et~al.} 2006, \apj, 640, 574

\end{thebibliography}

\end{document}